\documentclass[ aps, prd, reprint, nofootinbib, amsmath, amssymb, floatfix]{revtex4-2}

\usepackage{graphicx}
\usepackage[dvipsnames]{xcolor}
\usepackage{hyperref}
\hypersetup{colorlinks, urlcolor=BlueViolet, citecolor=Plum, linkcolor=PineGreen}

\usepackage{cleveref}

\usepackage{slashed}

\DeclareMathOperator{\Tr}{Tr}

\begin{document}

\title{New Limits on Light Dark Matter-Nucleon Scattering}

\author{Peter Cox}\email{peter.cox@unimelb.edu.au}
\author{Matthew J. Dolan}\email{matthew.dolan@unimelb.edu.au}
\author{Joshua Wood}\email{jwood5@student.unimelb.edu.au}

\affiliation{ARC Centre of Excellence for Dark Matter Particle Physics, \\
School of Physics, The University of Melbourne, Victoria 3010, Australia}

\begin{abstract}
     We derive new bounds on hadronically-interacting, sub-GeV mass dark matter. First, we show that one-loop interactions with photons can be sufficient to maintain equilibrium between the dark matter and Standard Model sectors at MeV temperatures, resulting in constraints from Big Bang Nucleosynthesis. Using chiral perturbation theory, we find that this leads to an upper bound on the dark-matter--nucleon scattering cross-section that is orders of magnitude stronger than existing astrophysical constraints. Furthermore, we show that even if these interactions remain out of equilibrium, there is an irreducible freeze-in abundance of dark matter that can easily overclose the universe. We also compute new bounds from rare Kaon decays that can provide even stronger constraints. Our results have significant implications for future direct detection experiments aiming to search for MeV-scale dark matter. 
\end{abstract}

\maketitle


\section{Introduction}

In the past decade, dark matter in the sub-GeV mass regime has been an exciting frontier in particle physics. The development of new models and mechanisms for achieving the observed relic density has in turn led to a wealth of proposals for direct detection experiments with very low thresholds and new accelerator-based experiments. 

It's well-known, however, that sub-GeV dark matter models face stringent constraints from both the early universe and precision measurements of the Standard Model (SM). In this work, we use a low-energy effective model to compute new bounds on hadronically-interacting dark matter from Big Bang Nucleosynthesis (BBN) and rare Kaon decays. The bounds we derive are orders of magnitude stronger than existing model-independent constraints for dark matter masses below $100\,$MeV, with significant implications for proposed direct detection experiments targeting this low-mass regime.

The success of the SM in accounting for the measured primordial abundances of the light elements severely restricts the density of additional degrees of freedom during BBN. The implications for light, thermal relic dark matter annihilating into either $e^\pm/\gamma$ or neutrinos have been studied in detail~\cite{Kolb:1986nf,Serpico:2004nm,Boehm:2012gr,Berezhiani:2012ru,Boehm:2013jpa,Nollett:2013pwa,Steigman:2014uqa,Nollett:2014lwa,Kawasaki:2015yya,Wilkinson:2016gsy,Escudero:2018mvt,Depta:2019lbe,Berlin:2019pbq,Sabti:2019mhn,Sabti:2021reh,Giovanetti:2021izc,Chu:2022xuh,An:2022sva}, leading to a bound on the dark matter mass of $m_\chi > 0.5$\,MeV~\cite{Sabti:2021reh}. An even stronger bound can be obtained using observations of the Cosmic Microwave Background (CMB), which constrain the energy density at the time of recombination, 
as parametrised by the effective number of neutrinos $N_\text{eff}$. 

Dark matter that interacts only hadronically is, at least naively, expected to be out-of-equilibrium during BBN due to the exponentially suppressed hadron abundance. Existing BBN bounds on hadronically-interacting dark matter have therefore been derived under the assumption that the dark matter was once in equilibrium at higher temperatures and its abundance persisted un-diluted during the subsequent cosmological evolution. For example, requiring that the dark matter decoupled prior to the QCD phase transition tightly constrains its interactions~\cite{Knapen:2017xzo,Krnjaic:2019dzc}.

There are, however, other processes that can bring the dark-matter into equilibrium with the SM thermal bath at BBN temperatures. For if dark matter interacts with hadrons, it inevitably interacts with photons via 1-loop diagrams involving pions and kaons.\footnote{Processes mediated via hadronic loops can also be relevant for direct detection, as recently discussed in Ref.~\cite{Diamond:2023fsm}.} Processes such as $\gamma\gamma\to\bar\chi\chi$ can then maintain equilibrium between the dark matter and the SM at low temperatures. The requirement that this process is out of equilibrium at $T\approx10$\,MeV can be translated into an upper bound on the dark-matter--nucleon cross-section. Importantly, such a bound is independent of the detailed thermal history at earlier times.

Interactions that are too weak to bring the dark matter into equilibrium with the SM can still result in substantial dark matter production via freeze-in at low temperatures. This irreducible abundance of dark matter, produced immediately prior to BBN, is very difficult to deplete without either contributing to $N_\text{eff}$ or diluting the baryons. We show that requiring that this population of dark matter does not overclose the universe imposes even stronger constraints on the dark-matter nucleon cross-section than BBN.

Rare meson decays to invisible final states provide another important constraint on hadronically-interacting dark matter. Decays such as $K \to \pi + \text{invisible}$ have previously been used to place stringent bounds on dark matter models with a light scalar mediator, where the mediator is produced on-shell in the decay. In this work, we consider low-energy effective models for light hadronically-interacting dark matter, assuming a contact interaction up to energies at least of $\mathcal{O}(m_K)$, and derive new bounds from the decay directly to dark matter, $K^+ \to \pi^+ \bar\chi \chi$.

The above considerations apply quite generically to hadronically-interacting dark matter, although the quantitative bounds are dependent on the form of the interaction. We study dark matter which couples either to gluons or quarks. We use chiral perturbation theory to calculate the dark-matter--photon scattering rate and meson decay widths, and derive bounds on the dark-matter--nucleon cross-section relevant for direct detection. Our bounds are many orders of magnitude stronger than existing cosmological/astrophysical bounds on dark-matter--baryon interactions from Milky Way sub-halos~\cite{Nadler:2019zrb,DES:2020fxi,Maamari:2020aqz,Buen-Abad:2021mvc}, Lyman-$\alpha$~\cite{Dvorkin:2013cea,Xu:2018efh,Rogers:2021byl}, and CMB anisotropies~\cite{Chen:2002yh,Gluscevic:2017ywp,Slatyer:2018aqg}. We discuss the implications of our results for future direct detection experiments targeting the MeV mass regime.


\section{Dark matter Interactions}
\label{sec:model}

We consider dark matter that interacts with the SM via an interaction of the general form
\begin{equation} \label{eq:interaction}
    \mathcal{L} \supset \mathcal{O}_\chi \mathcal{O}_\text{SM} \,,
\end{equation}
where $\mathcal{O}_\chi$ is an operator that is bilinear in the dark matter field $\chi$ and $\mathcal{O}_\text{SM}$ contains only SM fields. We consider the dark matter field to be either a scalar or fermion. In this work, we restrict these operators to be Lorentz scalars, leaving the generalisation for future work.

We focus on two particular choices for $\mathcal{O}_\text{SM}$ that are expected to arise in minimal UV completions:\\

\underline{Case 1}: Gluon-coupled
\begin{equation} \label{eq:O_G}
    \mathcal{O}_\text{SM}^G = \frac{\alpha_s}{8\pi} G^{a,\mu\nu}G_{a,\mu\nu} \,.
\end{equation}

\underline{Case 2}: Quark-coupled
\begin{equation} \label{eq:O_q}
    \mathcal{O}_\text{SM}^q = \sum_{q=u,d,s} m_q \bar{q} q + \frac{c_G\alpha_s}{8\pi} G^{a,\mu\nu}G_{a,\mu\nu} + \frac{c_\gamma\alpha}{8\pi} F^{\mu\nu} F_{\mu\nu} \,,
\end{equation}
with $c_G=-2$ and $c_\gamma=3$. The last two terms are obtained by integrating out the heavy quarks, assuming couplings of the form $m_q \bar{q} q$ in the UV.

Note that the quark-coupled case is closely related to models in which the Higgs acts as a portal between the SM and dark matter. In the latter case, one would instead have $c_\gamma=5/3$ as well as additional couplings to the light leptons.

We will be concerned with dark-matter--SM interactions at low energies, where the quarks and gluons are confined. We restrict ourselves to the case where the mediating particle(s) are sufficiently massive that the interactions with hadrons can be described using the local operators\footnote{Minimal models of hadronically-interacting dark matter with a sub-GeV scalar mediator have been explored in e.g.~\cite{Krnjaic:2015mbs,Green:2017ybv,Knapen:2017xzo}.}
\begin{equation} \label{eq:O_chi}
    \mathcal{O}_\chi = 
    \begin{cases}
        \frac{1}{\Lambda^2} \chi^* \chi & \text{(complex scalar)} \,, \\
        \frac{1}{\Lambda^3} \bar{\chi} \chi & \text{(Dirac fermion)} \,,
    \end{cases}
\end{equation}
with the scale $\Lambda$ parameterising the strength of the interaction between the dark matter and the SM. Note that we make no assumption about the form of the operator $\mathcal{O}_\chi$ describing the interaction with quarks and gluons at higher energies, which is in general non-local. 

The gluon-coupled and quark-coupled cases both give rise to spin-independent dark-matter--nucleon scattering at low energies. The cross-section can be calculated by matching onto heavy baryon chiral perturbation theory, following the approach of Ref.~\cite{Bishara:2016hek}. The result is
\begin{equation} \label{eq:sigma_chiN}
    \sigma_{\chi N} = \frac{1}{4\pi} \frac{|A_{\chi N}^X|^2}{\Lambda^4} 
    \begin{cases}
        1 & \text{(scalar)} \,, \\
        4m_\chi^2/\Lambda^2 & \text{(fermion)} \,,
    \end{cases}
\end{equation}
with
\begin{align}
    A^G_{\chi N} &= \frac{1}{9} m_G \,, \label{eq:AchiNG} \\
    A^q_{\chi N} &= \frac{1}{9} c_G m_G + \sum_{q=u,d,s} m_q b_0 \,, \label{eq:AchiNq}
\end{align}
where \cref{eq:AchiNG,eq:AchiNq} correspond to the gluon-coupled and quark-coupled cases, respectively. The low-energy constants $m_G=(847\pm8)\,$MeV and $b_0=-3.2\pm0.3$ are obtained from lattice data~\cite{FlavourLatticeAveragingGroupFLAG:2021npn}. Further details can be found in \cref{app:ChPT}.


\section{Dark-matter--photon Scattering \& Thermalisation}
\label{sec:DM-photon}

We want to determine whether the dark matter and SM sectors were in thermal equilibrium at $T\approx10\,$MeV. At this temperature, the only abundant SM degrees of freedom are $e^\pm$, $\nu$, and $\gamma$. In the models we consider, the dominant interaction of the dark matter is with the photons, since interactions with electrons and neutrinos are suppressed by at least an additional factor of $\alpha$.

There are three 2-to-2 processes involving dark matter and photons: $\chi \gamma \to \chi \gamma$, $\bar\chi \chi \to \gamma \gamma$ and $\gamma \gamma \to \bar\chi \chi$. The first two processes have dark matter in the initial state and their rates therefore depend on the dark matter density and momentum distribution. To avoid making an assumption about the initial dark matter distribution, we instead focus on the process $\gamma \gamma \to \bar\chi \chi$; if its rate is rapid compared to the Hubble expansion rate, then the dark matter will be brought into equilibrium, regardless of its initial distribution.

\begin{figure}
    \includegraphics[width=0.3\columnwidth]{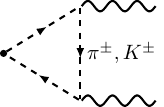}
    \hspace{2em}
    \includegraphics[width=0.3\columnwidth]{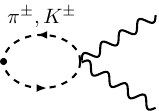}
    \caption{Feynman diagrams contributing to $\langle \gamma\gamma | \sum_q m_q \bar{q} q | 0 \rangle$ at leading order. There is an additional diagram obtained by interchanging the photons in the first diagram. \label{fig:qq-loop}}
\end{figure}

To calculate the rate of the process $\gamma \gamma \to \bar\chi \chi$ at temperatures below the QCD phase transition, we require the amplitude $\langle \gamma \gamma  | \mathcal{O}_\text{SM} | 0\rangle$ at scales below $\Lambda_{QCD}$. This is calculated by matching onto the $SU(3)_L \times SU(3)_R$ chiral Lagrangian. The leading-order contribution to $\langle \gamma\gamma | \sum_q m_q \bar{q} q | 0 \rangle$ is then given by the one-loop meson diagrams in \cref{fig:qq-loop} with pions and kaons in the loops. The amplitude $\langle \gamma\gamma  | \alpha_s GG | 0 \rangle$ can be obtained by relating it to $\langle \gamma\gamma  | T^\mu_\mu | 0 \rangle$, where $T^{\mu\nu}$ is the energy-momentum tensor~\cite{Leutwyler:1989tn}. In the quark-coupled case there is also a tree-level contribution from the direct coupling to photons. We simply quote the final results for the amplitudes here, with the details deferred to \cref{app:ChPT}:
\begin{equation} \label{eq:ME_OSM}
    \langle \gamma\gamma  | \mathcal{O}_\text{SM}^X | 0 \rangle = A^X_{\gamma\gamma}(s) \left(s\,g_{\mu\nu} -2k_{1\nu}k_{2\mu}\right) \epsilon_1^{\mu*} \epsilon_2^{\nu*} \,,
\end{equation}
with
\begin{align}
    A^G_{\gamma\gamma}(s) &= -\frac{\alpha}{36\pi} \bigg( 2 - \frac{1}{2} \sum_{i=\pi,K} \frac{(4 + x_i)}{x_i} F_0(x_i) \bigg) \,, \label{eq:A_G} \\
    A^q_{\gamma\gamma}(s) &= -\frac{\alpha}{4\pi} \bigg( c_\gamma + \frac{1}{2} \sum_{i=\pi,K} F_0(x_i) \bigg) + c_G A^G_{\gamma\gamma} \label{eq:A_q} \,.
\end{align}
The summations in the above expressions are over the charged mesons, with $x_i=4m_i^2/s$. The momenta and polarisation vectors of the photons are denoted by $k_{1,2}$ and $\epsilon_{1,2}$, respectively, and $s=(k_1+k_2)^2$. An explicit expression for the standard loop function $F_0(x)$ is given in the Appendix. Note that for $s\sim(10\,\text{MeV})^2$, we have $x_\pi,x_K \gg 1$ such that
\begin{equation} \label{eq:A_approx}
    A^G_{\gamma\gamma} \approx -\frac{7\alpha}{108\pi} \,, \qquad
    A^q_{\gamma\gamma} \approx \frac{131\alpha}{108\pi} \,.
\end{equation}

Computing the dark matter amplitude $\langle 0 | \mathcal{O}_\chi | \bar\chi \chi \rangle$ for the operators in \cref{eq:O_chi} is straightforward. For complex scalar dark matter, combining this with \cref{eq:A_G} or \eqref{eq:A_q} gives the polarisation-averaged cross-section
\begin{equation} \label{eq:sigma_photon}
    \sigma_{\gamma\gamma\to\bar\chi\chi}(s) = \frac{s}{32 \pi}\sqrt{1-\frac{4m_\chi^2}{s}} \frac{|A^X_{\gamma\gamma}(s)|^2}{\Lambda^4} \,.
\end{equation}
For Dirac fermion dark matter, the above result should be multiplied by $2(s-4m_\chi^2)/\Lambda^2$. 

The thermally averaged scattering rate, assuming Maxwell-Boltzmann statistics, is
\begin{align} 
    \langle\Gamma_{\gamma\gamma\to\bar\chi\chi}\rangle &= \frac{1}{n_\gamma} \frac{T}{16 \pi^4} \int_{4m_\chi^2}^\infty ds\, s^{3/2} \sigma_{\gamma\gamma\to\bar\chi\chi}(s) K_1\left(\frac{\sqrt{s}}{T}\right) \,, \label{eq:Gamma} \\
    &\simeq \frac{3T}{4\pi^3} \left(\frac{T}{\Lambda}\right)^4 |A^X_{\gamma\gamma}|^2 
    \begin{cases}
       1 & \text{(scalar)} \,, \\
       96 (T/\Lambda)^2 & \text{(fermion)} \,,
    \end{cases} \label{eq:Gamma_approx}
\end{align}
where the equilibrium photon distribution $n_\gamma = 2T^3/\pi^2$ and $K_1$ is a modified Bessel function of the second kind. The approximation in the second line assumes $m_\chi \ll T \ll m_\pi$, with $A^X_{\gamma\gamma}$ given by \cref{eq:A_approx}. Using \cref{eq:sigma_chiN}, this can be expressed in terms of the dark-matter--nucleon cross-section,
\begin{equation} \label{eq:Gamma_sigmaN}
    \langle\Gamma_{\gamma\gamma\to\bar\chi\chi}\rangle \simeq  \sigma_{\chi N} \frac{3T^5}{\pi^2} \frac{|A^X_{\gamma\gamma}|^2}{|A^X_{\chi N}|^2}
    \begin{cases}
       1 & \text{(scalar)} \,, \\
       24 (T/m_\chi)^2 & \text{(fermion)} \,.
    \end{cases}
\end{equation}

\Cref{eq:Gamma_sigmaN} clearly illustrates that a sufficiently large dark-matter--nucleon cross-section inevitably leads to rapid dark matter interactions with photons. These interactions will bring the dark matter into equilibrium with the SM bath if the above rate is faster than the Hubble rate $H(T) \simeq 1.66 T^2\, g_*^{1/2}(T)$, with g$_*(10\,\text{MeV}) \approx 10.75$. Note that in our numerical results in \cref{sec:results} we use the full expression for $\Gamma_{\gamma\gamma\to\bar\chi\chi}$ in \cref{eq:Gamma}.


\section{Constraints from BBN}
\label{sec:BBN-CMB}

The presence of additional relativistic degrees of freedom modifies the abundances of the light elements produced during BBN. Of particular importance are helium ($Y_P$) and deuterium ($D/H|_P$) for which the primordial abundances can be inferred with a high degree of precision from observations~\cite{ParticleDataGroup:2022pth}.

Assuming that the dark matter was in equilibrium prior to neutrino decoupling, its impact on BBN depends on when it decouples from the photons. We therefore separate our discussion into three regimes:

\begin{enumerate}
\item \emph{Dark matter decouples while relativistic, \\
before $e^\pm$ annihilation} \\
In this case, the only effect of the dark matter is to increase the expansion rate, which can be parametrised as a contribution to the effective number of relativistic degrees of freedom,
\begin{equation}
    N_\text{eff} = \frac{8}{7} \left(\frac{11}{4}\right)^{4/3} \frac{\rho_\nu + \rho_\chi}{\rho_\gamma} \,.
\end{equation}
The increase in $N_\text{eff}$ causes weak and nuclear reaction rates to freeze-out earlier, leading to an increase in both $Y_P$ and $D/H|_P$~\cite{Steigman:1977kc}. Comparing the resulting BBN predictions with observations, Ref.~\cite{Yeh:2022heq} obtained an upper bound of $\Delta N^\text{BBN}_\text{eff} < 0.413$ (95\%\,CL). This is sufficient to exclude a single real scalar in equilibrium with the photons, which contributes $\Delta N_\text{eff}=4/7\approx0.57$.

Dark matter that decouples while relativistic will, of course, overclose the universe. We defer a more detailed discussion of the dark matter abundance to \cref{sec:abundance}.

\item \emph{Dark matter decouples while relativistic, \\
after $e^\pm$ annihilation} \\
As in the previous case, the relativistic dark matter directly increases the expansion rate during BBN. In addition, the entropy released by $e^\pm$ annihilation is shared between the dark matter and the photons. This leads to a larger neutrino-to-photon temperature ratio than in the SM,
\begin{equation}
    \left(\frac{T_\nu}{T_\gamma}\right)^3 = \frac{4 + 2g_*^\chi}{11 + 2g_*^\chi} \,,
\end{equation}
where $g_*^\chi = 2$ or $7/2$ for complex scalar or Dirac fermion dark matter, respectively. The increase in $T_\nu/T_\gamma$ has the effect of further increasing the expansion rate for a given photon temperature, relative to the SM. 

As far as we are aware, no dedicated BBN analysis has been performed for this type of scenario. Nevertheless, it's clear that the increased expansion rate after $e^\pm$ annihilation will further increase both $Y_P$ and $D/H|_P$ relative to case 1 above. This scenario is therefore also excluded. The dark matter will also overclose the universe.

\item \emph{Dark matter decouples when non-relativistic} \\
In this case, the dark matter transfers its entropy to the photons when it decouples. This decreases $T_\nu/T_\gamma$ and therefore decreases the expansion rate for a given photon temperature. It also has the effect of diluting the baryons relative to the photons. For dark matter masses $m_\chi\lesssim2$\,MeV, the net effect is an increase in $Y_P$~\cite{Serpico:2004nm}. Ref.~\cite{Sabti:2021reh} was consequently able to set a 95\%\,CL bound of $m_\chi>0.5\,$MeV ($m_\chi>0.7\,$MeV) on the mass of an additional complex scalar (Dirac fermion) in equilibrium with the photons during BBN.
\end{enumerate}

Combining the three regimes above\footnote{We have not discussed the case of semi-relativistic decoupling; however, we expect this to also be excluded for dark matter masses below roughly an MeV.}, the overall conclusion is that dark matter with mass $m_\chi\lesssim 0.5\,\mathrm{MeV}$ is excluded, if it was in equilibrium with the photons prior to or during BBN. Note that in deriving these bounds we did not need to assume that $\chi$ saturates the dark matter abundance, and these bounds are applicable to any sub-MeV mass particle in equilibrium with the photons.

Finally, we note that the BBN constraints above can be strengthened in a combined fit with CMB observations. There are two important effects. First, the CMB provides an independent measurement of the baryon abundance. Second, it tightly constrains $N_\text{eff}$ at recombination, $N^\text{CMB}_\text{eff}=2.99\pm0.17$~\cite{Planck:2018vyg}. The CMB measurement of the dark matter density leads to even stronger constraints, which we discuss in \cref{sec:abundance}.


\section{Meson decays}
\label{sec:meson}

Rare meson decays can also provide strong constraints on hadronically-interacting dark matter. These bounds have been well studied within the context of specific models, in particular those with a singlet scalar that mixes with the Higgs~\cite{Krnjaic:2015mbs,Bird:2004ts}. Here, our aim is to remain as model-independent as possible and we derive new bounds within the low-energy effective theory given by \cref{eq:interaction,eq:O_G,eq:O_q,eq:O_chi}.

We focus on the bounds from $K$-meson decays, assuming that the contact interaction in \cref{eq:O_chi} remains valid up to energies of $\mathcal{O}(m_K)$. (While similar bounds could also be derived from $B$ decays, these would require additional assumptions about the form of the interaction at higher energies.) The measurement of the SM decay $\text{BR}(K^+ \to \pi^+ \bar\nu \nu)=(1.06\pm0.4)\times 10^{-10}$~\cite{NA62:2021zjw} places a strong upper bound on the branching ratio to other invisible final states, including dark matter.

Within the low-energy effective Lagrangians given by \cref{eq:interaction,eq:O_G,eq:O_q,eq:O_chi}, the $s \to d$ transition is mediated purely by the SM electroweak interactions. The $K^+ \to \pi^+ \bar\chi \chi$ decay rate can then be calculated in chiral perturbation theory using the low-energy effective weak Lagrangian. The details of this calculation are provided in \cref{app:meson-decay}. The resulting differential decay rate is
\begin{multline} \label{eq:Gamma_K_IR}
    \frac{d\Gamma}{dq^2} = \frac{1}{256\pi^3 m_K^3} |\mathcal{M}(q^2)|^2 \sqrt{1-\frac{4m_\chi^2}{q^2}}\, \lambda(m_K^2,m_\pi^2,q^2) \,,
\end{multline}
where $q^2$ is the invariant mass of the pair of dark matter particles and $\lambda(a,b,c)$ is the K\"all\'en function. For complex scalar dark matter, the matrix element is 
\begin{equation} \label{eq:Kpi_ME_IR}
   \mathcal{M}(q^2) = \sqrt{2} G_F V_{ud} V_{us}^* g_8 f^2 \frac{c_G}{9\Lambda^2} (m_K^2 + m_\pi^2 - q^2) \,,
\end{equation}
where $G_F$ is the Fermi constant, $V_{ij}$ are CKM matrix elements, and the low-energy constant $g_8=3.07\pm0.14$~\cite{Pich:2021yll}. The gluon-coupled and quark-coupled cases correspond to $c_G=1$ or $c_G=-2$, respectively. For Dirac fermion dark matter, the amplitude-squared contains an additional factor of $4(q^2-3m_\chi^2)/\Lambda^2$. 

There can of course be additional contributions to the $K^+ \to \pi^+ \bar\chi \chi$ decay rate that are generated within the UV completions of the effective low-energy models we consider. While we aim to remain as model-agnostic as possible, in the quark-coupled case there is a UV contribution that arises quite generically and does not require the specification of a full model. We assume that $\mathcal{O}_\text{SM}^q=\sum_q m_q \bar{q} q$ in the UV, with direct couplings to the heavy quarks. There is then an additional contribution to the decay that arises from integrating out the top-quark at one-loop. The resulting matrix element is given by (see \cref{app:meson-decay})
\begin{equation} \label{eq:Kpi_ME_UV}
    \mathcal{M}_{UV}^q = -\frac{\sqrt{2} G_F m_t^2 V_{td} V_{ts}^*}{16\pi^2} \frac{m_K^2}{2\Lambda^2}F_t(m_W^2/m_t^2)\,,
\end{equation}
with $F_t(m_W^2/m_t^2) \approx 0.6$. This UV contribution to the decay rate dominates over that in \cref{eq:Kpi_ME_IR}.

We make the simplifying approximation that the signal acceptance at NA62 is the same for the decays $K^+\to\pi^+\bar\nu\nu$ and $K^+\to\pi^+\bar\chi\chi$. Requiring consistency with the measurement at $2\sigma$ then leads to the bound $\text{BR}(K^+\to\pi^+\bar\chi\chi) < 10^{-10}$. Note, however, that the $q^2$-distributions in the gluon-coupled and quark-coupled cases differ from the SM decay and each other. The effect on the signal acceptance is expected to be small, but should be taken into account to obtain a more precise bound. Furthermore, the different dependence on $q^2$ could ultimately provide a means to distinguish between the different decay modes.

The upper limit on the $K^+ \to \pi^+ \bar\chi \chi$ branching ratio can be translated into a bound on the dark-matter--nucleon cross-section by substituting for $\Lambda$ using \cref{eq:sigma_chiN}. The resulting constraints are, however, significantly more model-dependent than those from BBN; we shall see this explicitly in our numerical results in the following section.


\section{Irreducible Dark Matter Abundance}
\label{sec:abundance}

The constraints we have discussed so far, from BBN and meson decays, apply independently of whether or not $\chi$ is the dark matter. We now turn our attention to the abundance of $\chi$. Our goal is to derive constraints that are independent of both the early cosmological history and the form of the dark matter interactions at energies above $\mathcal{O}(100)$\,MeV. We therefore only require that the interactions of $\chi$ at low temperatures do not lead to an overabundance of dark matter. As we shall see, this already results in very strong bounds. 

First, consider the case where $\chi$ is out of equilibrium with the SM at low temperatures (and so unconstrained by BBN). Regardless of whether $\chi$ was once in equilibrium at earlier times, there is an \emph{irreducible} abundance of dark matter produced via the freeze-in process $\gamma\gamma\to\bar\chi\chi$. 

The dark matter yield $Y_\chi$ is obtained by integrating the Boltzmann equation
\begin{equation}
    \label{eq:BE_yield}
    \frac{dY_{\chi}}{dT}=-\frac{2\langle \sigma v\rangle_{\gamma\gamma\rightarrow\chi\chi} n_{\gamma}^2}{HsT}\bigg(1+\frac{T}{3}\frac{d\ln g_{*s}}{dT}\bigg) \,,
\end{equation}
where $g_{*s}(T)$ is the number of relativistic entropic degrees of freedom and $\langle \sigma v\rangle_{\gamma\gamma\rightarrow\chi\chi} = \langle\Gamma_{\gamma\gamma\to\bar\chi\chi}\rangle/n_\gamma$, with $\langle\Gamma_{\gamma\gamma\to\bar\chi\chi}\rangle$ given in \cref{eq:Gamma}. This freeze-in process is UV-dominated and the dark matter yield increases with the maximum temperature. For viable BBN, the universe must have reheated to a temperature of at least a few MeV~\cite{deSalas:2015glj,Hasegawa:2019jsa}. We calculate the irreducible abundance of dark matter produced below $T=10\,$MeV and use this to constrain the strength of the dark matter interactions.

Next, consider the alternative case where the process $\gamma\gamma\to\bar\chi\chi$ is sufficient to bring the DM into equilibrium with the SM at a temperature $T < \Lambda_{QCD}$. If the dark matter subsequently decouples while relativistic it will overclose the universe. If, on the other hand, it undergoes non-relativistic freeze-out, then the $\bar\chi\chi\to\gamma\gamma$ annihilation cross-section should be at least as large as the thermal relic cross-section, $\langle\sigma v\rangle \gtrsim \text{few} \times 10^{-26}\,\text{cm}^3/\text{s}$, to avoid over-production of dark matter.

In all cases, there is a population of dark matter that is produced either just prior to and/or during BBN. If this abundance is too large it could, at least in principle, be depleted prior to recombination. This might be achieved if the dark matter can annihilate into dark radiation, is able to decay, or if there is additional entropy production that dilutes the dark matter. In practice, however, significant depletion is difficult to achieve without either conflicting with the CMB constraint on $N_\text{eff}$ or diluting the baryons.


\section{Results \& Discussion}
\label{sec:results}

\begin{figure*}
    \includegraphics[width=0.49\textwidth]{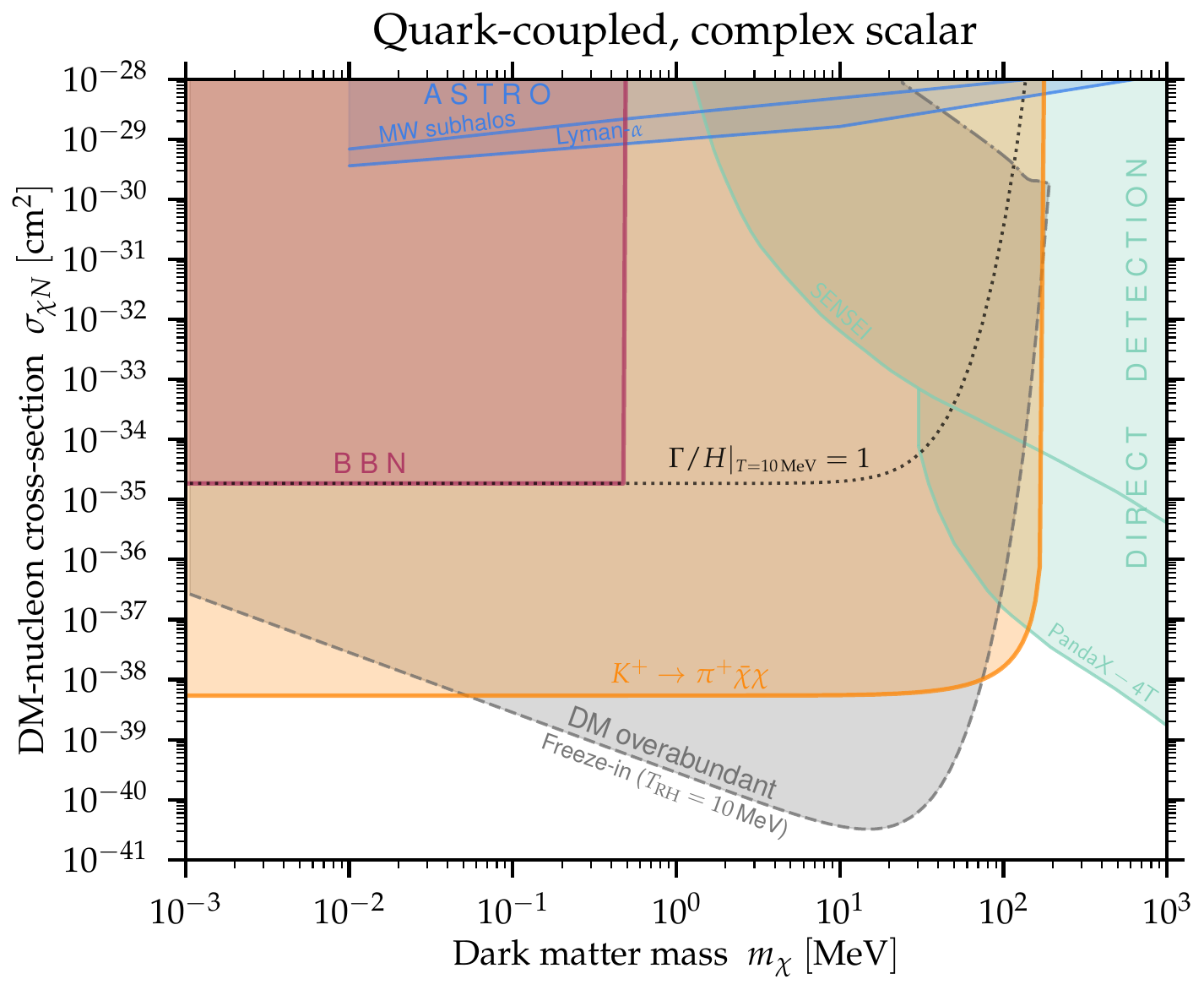}
    \includegraphics[width=0.49\textwidth]{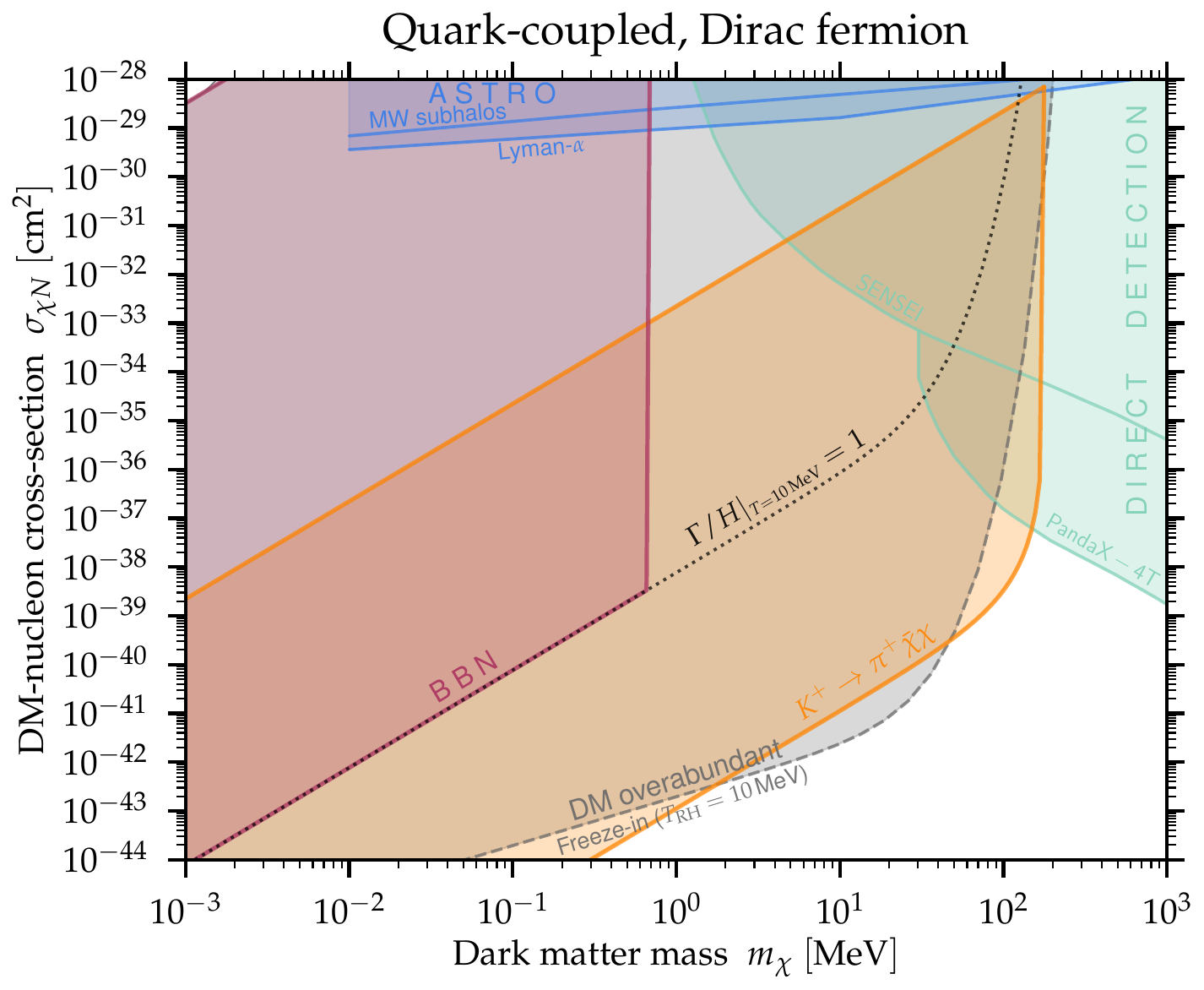} \\
    \vspace{2ex}
    \includegraphics[width=0.49\textwidth]{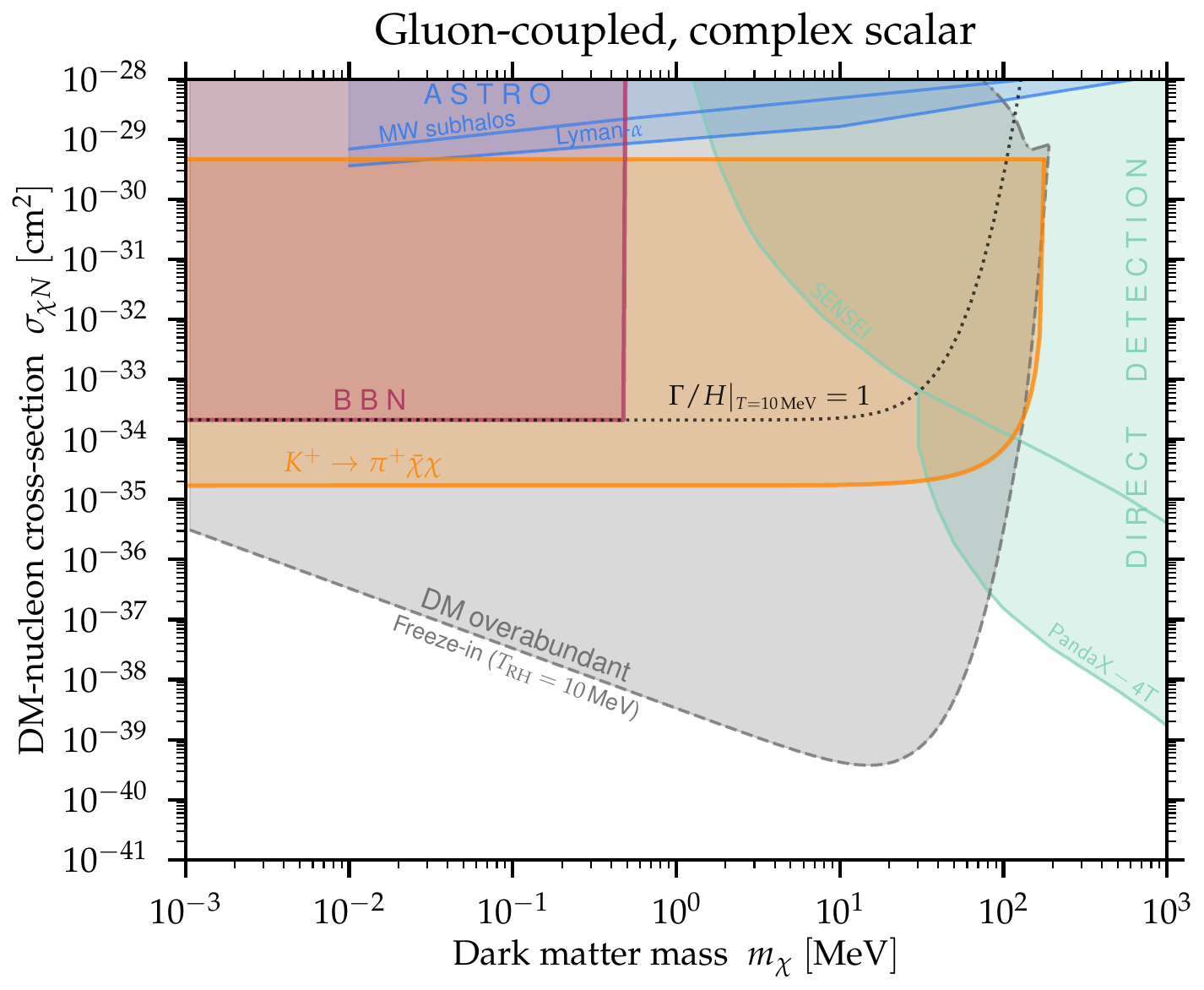}
    \includegraphics[width=0.49\textwidth]{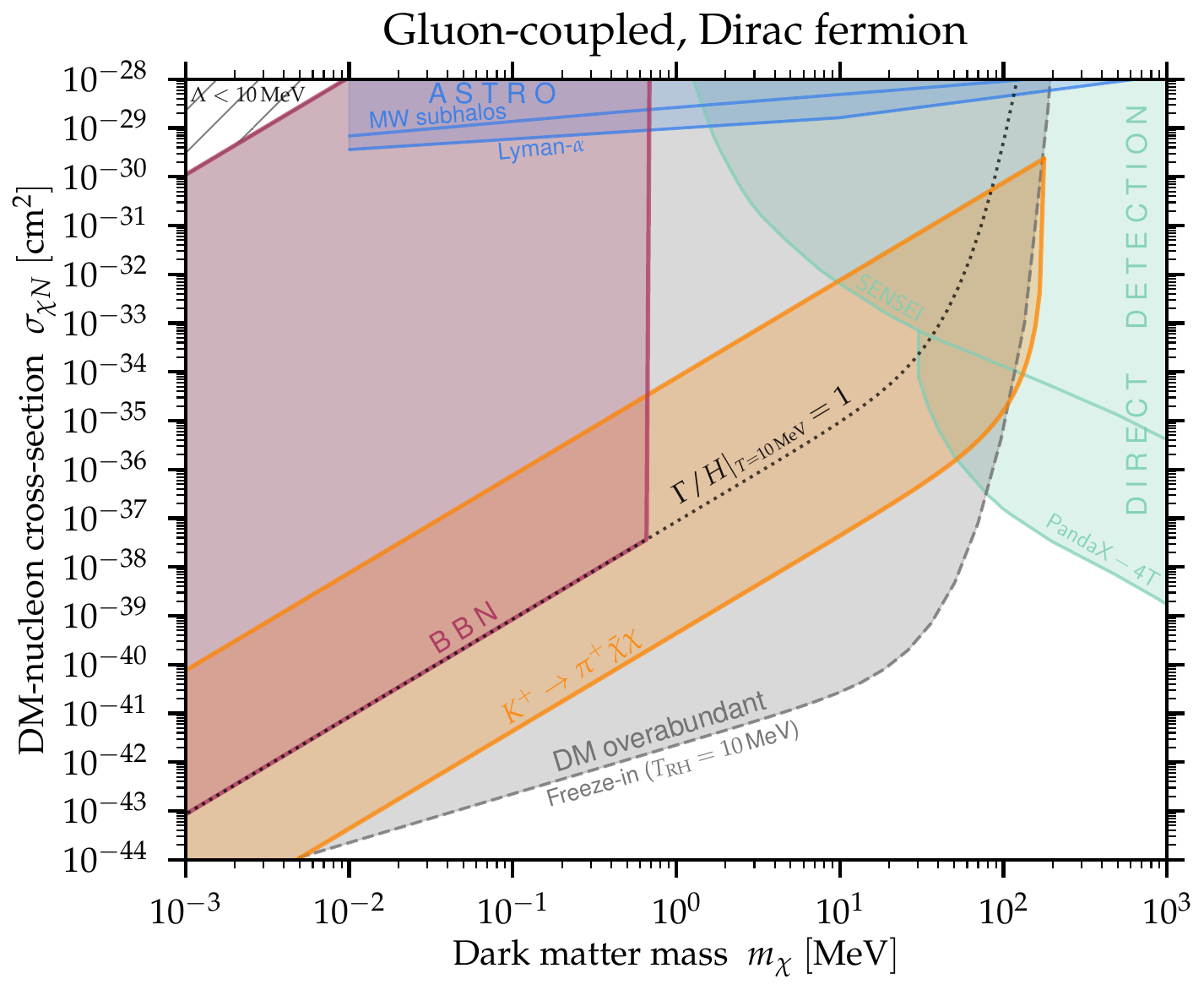}
    \caption{Constraints on the dark-matter--nucleon cross-section. The red shaded regions are excluded by BBN, requiring that the dark matter is out of equilibrium with the SM at $T=10$\,MeV. The orange region is excluded by $K^+\to\pi^+\bar\chi\chi$. The grey shaded regions are excluded by dark matter overabundance, with the dashed grey lines denoting where the low-temperature freeze-in process $\gamma\gamma\to\bar\chi\chi$ saturates the observed relic abundance, assuming instantaneous reheating with $T_\text{RH}=10\,$MeV. Along the grey dot-dashed lines the observed relic abundance is obtained via freeze-out. The blue and green regions are excluded by existing bounds from structure formation~\cite{Rogers:2021byl,Buen-Abad:2021mvc} and direct detection experiments~\cite{SENSEI:2023zdf,PandaX:2023xgl}, respectively. \label{fig:limits}}
\end{figure*}

Based on the discussion in \cref{sec:BBN-CMB}, sub-MeV dark matter that was in equilibrium with the SM in the early universe is excluded by BBN observations. We use this to place an upper bound on the dark-matter--nucleon scattering cross-section, $\sigma_{\chi N}$. As discussed in \cref{sec:DM-photon}, the rate of the process $\gamma\gamma\to\bar\chi\chi$ is proportional to $\sigma_{\chi N}$; hence, for sufficiently large cross-sections, the dark matter is brought into equilibrium with the photons, even at temperatures $T \ll \Lambda_{QCD}$ where the baryon abundance is highly suppressed.

For dark matter that interacts via a contact interaction, the ratio $\Gamma_{\gamma\gamma\to\bar\chi\chi}/H$ increases with temperature; hence, the upper bound on $\sigma_{\chi N}$ depends on the maximum temperature achieved in the early universe. Our aim is to set a conservative bound that does not require additional assumptions about the very early cosmological history. We therefore require that the dark matter remain out of equilibrium at a temperature of 10\,MeV, slightly above neutrino decoupling. Increasing this temperature will lead to a stronger bound on $\sigma_{\chi N}$, as can be seen from \cref{eq:Gamma_sigmaN}. Specifically, we impose $\Gamma_{\gamma\gamma\to\bar{\chi}\chi}/H < 1$ at $T=10\,$MeV. Note, however, that for values of $\Gamma_{\gamma\gamma\to\bar{\chi}\chi}$ slightly below this bound, there may still be sufficient production of dark matter to be in conflict with the BBN constraint on additional relativistic degrees of freedom. In this sense our bound is again conservative.

Our constraints on the dark-matter--proton cross-section derived from BBN are shown by the red excluded region in \cref{fig:limits}. The upper (lower) row is for the quark-coupled (gluon-coupled) case and the left (right) column is for complex scalar (Dirac fermion) dark matter. For a real scalar or Majorana fermion, the corresponding limits are weaker by approximately a factor of two. The different dependence on $m_\chi$ in the scalar and fermion cases can be understood from the fact that, for fixed $\sigma_{\chi N}$ and in the limit $m_\chi \ll T$, $\Gamma_{\gamma\gamma\to\bar\chi\chi}$ is independent of $m_\chi$ in the scalar case, while for fermionic dark matter it is proportional to $(T/m_\chi)^2$ (again, see \cref{eq:Gamma_sigmaN}).

The orange exclusion regions in \cref{fig:limits} show our bounds from $K^+\to\pi^+\bar\chi\chi$. In the gluon-coupled case, this bound is slightly stronger than that from BBN. In the quark-coupled case, the meson decay bound is significantly stronger due to the additional UV contribution to the decay width discussed in \cref{sec:meson}. Without this additional contribution (e.g. if there was no coupling to the top quark in the UV), the bounds would be comparable to the gluon-coupled case. The different $m_\chi$-dependence between scalar and fermionic dark matter again originates in the dark-matter--nucleon cross-section.

Notice that the meson decay bounds have a ceiling and do not extend to large cross-sections, especially for fermionic dark matter. The reason is that in this region the effective contact interaction description in \cref{eq:O_chi} is not valid for $K$ decays, since the corresponding value of $\Lambda$ is smaller than $m_K$. To achieve such large cross-sections therefore requires the particle(s) mediating the interaction between the dark matter and the SM to also have sub-GeV masses. There will generally then be bounds from meson decays in which the mediator is produced on-shell; however, these depend on the details of the UV-completion. Similarly, we do not consider bounds from $B$-meson decays or colliders, which are sensitive to the structure of the interaction at higher energies, and therefore model-dependent. 

In the region below the BBN bounds, where the dark matter remains out of equilibrium, there is an irreducible freeze-in abundance of dark matter produced via $\gamma\gamma\to\bar\chi\chi$, as discussed in \cref{sec:abundance}. The grey lines in \cref{fig:limits} show where this yields the observed dark matter abundance, assuming instantaneous reheating with a reheat temperature of $T_\text{RH}=10\,$MeV. The grey shaded regions are then strongly disfavoured due to overproduction of dark matter. For the gluon-coupled scenario, this gives the strongest constraint on $\sigma_{\chi p}$ across most of the dark matter mass range. 

In the regions below the dashed grey lines, either a higher reheat temperature (which leads to stronger BBN bounds) or other non-thermal production mechanisms would be required to obtain the correct abundance.\footnote{This could be achieved, for example, with non-instantaneous reheating if the maximum temperature during reheating is considerably higher than $T_\text{RH}=10\,$MeV; the additional freeze-in process $\pi\pi\to\bar\chi\chi$ can then contribute significantly to the dark matter abundance~\cite{Bhattiprolu:2022sdd}.} Finally, the grey dot-dashed lines show where the correct dark matter abundance is obtained via freeze-out. (For fermionic dark matter this occurs for larger cross-sections than shown in the figures.)

Our bounds are orders of magnitude stronger than previous cosmological constraints on $\sigma_{\chi N}$ coming from the effect of dark-matter--baryon interactions on the matter power spectrum as probed by Lyman-$\alpha$~\cite{Rogers:2021byl} or Milky Way sub-halos~\cite{Buen-Abad:2021mvc}. These previous bounds are shown in blue in \cref{fig:limits}. Lyman-$\alpha$ constraints on warm dark matter also disfavour dark matter masses below approximately 5\,keV~\cite{Ballesteros:2020adh}. At higher masses, the leading constraints from direct detection experiments~\cite{SENSEI:2023zdf,PandaX:2023xgl}, which assume that $\chi$ constitutes all of the dark matter, are shown in green.

Our results have significant implications for future direct detection experiments targeting sub-GeV mass hadronically-interacting dark matter. To reach parameter space that is not already constrained by BBN and meson decays, these experiments will require sensitivity to cross-sections at least as small as $10^{-35}\,\text{cm}^2$ (and significantly smaller again for fermionic dark matter). While such cross-sections are within the projected reach of future experiments for $\mathcal{O}(10\,\text{MeV})$ masses and above (see~\cite{Hertel:2018aal,Oscura:2022vmi,vonKrosigk:2022vnf,NEWS-G:2023qwh}), they may be challenging to reach in the sub-MeV regime.

Finally, we comment briefly on models of hadronically-interacting dark matter that do not fall into either the gluon-coupled or quark-coupled scenarios. For example, dark matter could interact only with first generation quarks, as in Refs.~\cite{Batell:2017kty,Batell:2018fqo}. (Although UV-completions will likely be challenged by flavour bounds.) We have verified that the bounds on such a scenario from BBN and meson decays are qualitatively similar to those presented in \cref{fig:limits}. It would be interesting to perform a detailed analysis for other Lorentz structures in the future.


\begin{acknowledgments}
We would like to thank Felix Kahlh\"ofer for comments on a draft version of the manuscript. This work was supported in part by the Australian Research Council through the ARC Centre of Excellence for Dark Matter Particle Physics CE200100008 and the Australian Government Research Training Program Scholarship initiative. P.C. is supported by the Australian Research Council Discovery Early Career Researcher Award DE210100446.
\end{acknowledgments}


\appendix

\section{}
\label{app:ChPT}

In this appendix, we compute the matrix elements for $\gamma\gamma\to\bar{\chi}\chi$ and $\chi N \to \chi N$ using chiral perturbation theory.


\subsection{Matching to the chiral Lagrangian}

To describe the dark-matter--SM interactions below the QCD scale, we match \cref{eq:interaction,eq:O_G,eq:O_q} on to the $SU(3)$ chiral Lagrangian.

The eight (pseudo-) Nambu Goldstone bosons parameterise the $(SU(3)_L \times SU(3)_R) / SU(3)_V$ coset space and can be described by the unitary matrix
\begin{equation}
    U = \exp(i\sqrt{2}\,\Pi/f) \,,
\end{equation}
with
\begin{equation}
    \Pi = 
    \begin{pmatrix}
        \frac{\pi^0}{\sqrt{2}} + \frac{\eta_8}{\sqrt{6}} & \pi^+ & K^+ \\
        \pi^- & -\frac{\pi^0}{\sqrt{2}} + \frac{\eta_8}{\sqrt{6}} & K^0 \\
        K^- & \bar{K}^0 & \frac{-2\eta_8}{\sqrt{6}}
    \end{pmatrix}
    \,,
\end{equation}
and where $f\approx92\,$MeV is the pion decay constant at leading order. Under chiral transformations, $U$ transforms linearly: $U \to V_R U V_L^\dagger$, with $V_{L(R)}$ the $SU(3)_{L(R)}$ transformation matrices.

The dark matter can be treated as an external source when considering QCD interactions. Above the QCD scale, the relevant Lagrangian can be written as
\begin{equation} \label{eq:QCD}
    \mathcal{L} = \mathcal{L}^0_\text{QCD} + s_G(x) \frac{\alpha_s}{12\pi} G^{a,\mu\nu}G_{a,\mu\nu} - \bar{q} s(x) q \,,
\end{equation}
where $\mathcal{L}^0_\text{QCD}$ is the massless QCD Lagrangian, $q=(u,d,s)$ and $s_G(x)$, $s(x)$ are local sources. This Lagrangian is invariant under $SU(3)_L \times SU(3)_R$ transformations, provided the sources are treated as spurions that transform as
\begin{equation}
    s(x) \to V_R s(x) V_L^\dagger\,, \qquad s_G(x) \to s_G(x) \,.
\end{equation}

The source $s(x)$ accounts for both the quark masses and the dark matter, and it is convenient to define $s=M_q( 1 + s_\chi)$. For complex scalar dark matter, comparing with \cref{eq:O_G,eq:O_q,eq:O_chi} gives

\underline{Case 1}: Gluon-coupled
\begin{align} \label{eq:sources_G}
    s_\chi = 0 \,, \quad s_G = \frac{3}{2\Lambda^2} \chi^* \chi \,.
\end{align}

\underline{Case 2}: Quark-coupled
\begin{align} \label{eq:sources_q}
    s_\chi = \frac{-1}{\Lambda^2} \chi^* \chi \,, \quad s_G = \frac{3c_G}{2\Lambda^2} \chi^* \chi \,.
\end{align}
For Dirac fermion dark matter one makes the replacement $\chi^* \chi/\Lambda^2 \to \bar{\chi}\chi/\Lambda^3$.

The Lagrangian \eqref{eq:QCD} can be matched on to the $SU(3)$ chiral Lagrangian at low energies. At $\mathcal{O}(p^2)$ in the derivative expansion this leads to
\begin{multline} \label{eq:L_ChPT}
    \mathcal{L}^\text{LO}_\text{ChPT} = \frac{f^2}{4} \left(1 + \frac{4}{27}s_G \right) \Tr[D^\mu U^\dagger D_\mu U] \\
    + \frac{B_0 f^2}{2} \left( 1 + s_\chi + \frac{2}{9} s_G \right) \Tr[M_q (U + U^\dagger )] \,,
\end{multline}
with the covariant derivative $D_\mu U = \partial_\mu U + i e A_\mu (Q_q U + U Q_q)$, where $Q_q=\mathrm{diag}(2/3,-1/3,-1/3)$. The coupling of the source $s$ is fixed by the $SU(3)_L \times SU(3)_R$ invariance of the Lagrangian. This is not the case for $s_G$, which is invariant under chiral transformations. To determine the operator it couples to, one requires that the trace of the energy-momentum tensor (i.e. the anomalous Ward identity of scale transformations) in the chiral effective theory is consistent with that of QCD~\cite{Voloshin:1980zf,Chivukula:1989ds,Leutwyler:1989tn,Donoghue:1990xh}. Expanding \eqref{eq:L_ChPT} to quadratic order in the charged meson fields yields
\begin{multline} \label{eq:L_ChPT2}
    \mathcal{L}^\text{LO}_\text{ChPT} \supset (1 + \frac{4}{27} s_G) \left( (D^\mu\pi^+)(D_\mu\pi^-) + (D^\mu K^+)(D_\mu K^-) \right) \\
    + (1 + s_\chi + \frac{2}{9} s_G) \left(m_\pi^2 \pi^+ \pi^- + m_K^2 K^+ K^- \right) \,,
\end{multline}
where we have used that the low-energy constant $B_0$ satisfies $B_0(m_u+m_d) \simeq m_{\pi^\pm}^2$ at leading order and neglecting QED corrections.


\subsection{\texorpdfstring{$\gamma \gamma \to \bar{\chi}\chi$}{y + y -> chi* + chi}}

Using the Lagrangian \eqref{eq:L_ChPT2}, we can compute the matrix elements needed for the process $\gamma\gamma\to\bar{\chi}\chi$. The relevant SM matrix elements have previously been computed in the context of light Higgs decays to photons~\cite{Leutwyler:1989tn}; we include the details here for completeness. 

We begin with $\langle \gamma\gamma | \bar{q} M_q q | 0 \rangle$, which at leading order is given by the one-loop meson diagrams in \cref{fig:qq-loop}. The result is 
\begin{equation} \label{eq:ME_qq}
    \langle \gamma\gamma | \bar{q} M_q q | 0 \rangle = - \frac{\alpha}{8\pi} \left(s\,g_{\mu\nu} -2k_{1\nu}k_{2\mu}\right) \epsilon_1^{\mu*} \epsilon_2^{\nu*} \sum_{i=\pi,K} F_0(x_i) \,,
\end{equation}
where $k_{i\mu}$ and $\epsilon_i^\mu$ are the momenta and polarisation vectors of the photons and $x_i=4m_i^2/s$, with $s=(k_1+k_2)^2$. The standard scalar loop function is
\begin{equation}
    F_0(x) = x \left(1 -xF(x) \right) \,, \label{eq:F0}
\end{equation}
with
\begin{equation}
    F(x) = - \frac{1}{4} \left[ \ln \left( 1 - \frac{2}{x} + \frac{2\sqrt{1-x}}{|x|} \right) \right]^2 \,,
\end{equation}
and $\lim_{x\to\infty}F_0(x) \to -1/3$.

\begin{figure}
    \includegraphics[width=0.4\columnwidth]{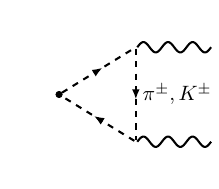}
    \hspace{2em}
    \includegraphics[width=0.4\columnwidth]{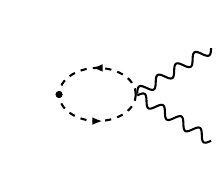} \\
    \hspace{2em}
    \includegraphics[width=0.4\columnwidth]{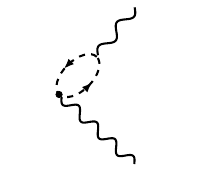}
    \hspace{2em}
    \includegraphics[width=0.4\columnwidth]{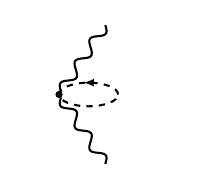}
    \caption{Feynman diagrams contributing to $\langle \gamma\gamma | \alpha_s GG | 0 \rangle$ at leading order, where the loops contain $\pi^\pm$ or $K^\pm$. There are two additional diagrams obtained by interchanging the photons in the diagrams of the first column. \label{fig:GG-loop}}
\end{figure}

We now move to the computation of the gluonic matrix element $\langle \gamma\gamma | \alpha_s GG | 0 \rangle$. The same two charged meson diagrams (\cref{fig:GG-loop} upper row) appear in this case; however, there are additional diagrams (\cref{fig:GG-loop} lower row) that originate in the first line of \cref{eq:L_ChPT2}. Together these give
\begin{multline} \label{eq:ME_GG}
    \langle \gamma\gamma  | \alpha_s GG | 0 \rangle = -\frac{2\alpha}{9} \bigg( 2 - \frac{1}{2} \sum_{i=\pi,K} \frac{(4 + x_i)}{x_i} F_0(x_i) \bigg) \\
    \times \left(s\,g_{\mu\nu} -2k_{1\nu}k_{2\mu}\right) \epsilon_1^{\mu*} \epsilon_2^{\nu*} \,.
\end{multline}

Finally, there is the tree-level contribution,
\begin{equation} \label{eq:ME_FF}
    \langle \gamma\gamma | \alpha F F | 0 \rangle = -2\alpha \left(s\,g_{\mu\nu} -2k_{1\nu}k_{2\mu}\right) \epsilon_1^{\mu*} \epsilon_2^{\nu*} \,.
\end{equation}

Combining the expressions in \cref{eq:ME_qq,eq:ME_GG,eq:ME_FF} with the appropriate constant coefficients, one obtains the matrix elements $\langle \gamma\gamma  | \mathcal{O}_\text{SM}^G | 0 \rangle$ and $\langle \gamma\gamma  | \mathcal{O}_\text{SM}^q | 0 \rangle$ given in \cref{eq:ME_OSM,eq:A_G,eq:A_q}.


\subsection{\texorpdfstring{$\chi N \to \chi N$}{chi + N -> chi + N}}

To describe dark-matter--nucelon scattering at low energies, we follow the approach of Ref.~\cite{Bishara:2016hek} which uses Heavy Baryon Chiral Perturbation Theory~\cite{Jenkins:1990jv}. The baryon momentum is split according to $p_\mu = m_N v_\mu + k_\mu$, with $m_N$ the baryon mass, $v_\mu$ the baryon four-velocity, and $k_\mu$ a soft momentum ($k\cdot v \ll m_N$). The baryon velocity is factored out of the low-energy dynamics by defining the field
\begin{equation}
    B_v = \exp(im_N\slashed{v} v_\mu x^\mu) B(x) \,,
\end{equation}
with the baryon octet matrix
\begin{equation}
    B = 
    \begin{pmatrix}
        \frac{\Sigma^0}{\sqrt{2}} + \frac{\Lambda}{\sqrt{6}} & \Sigma^+ & p \\
        \Sigma^- & -\frac{\Sigma^0}{\sqrt{2}} + \frac{\Lambda}{\sqrt{6}} & n \\
        \Xi^- & \Xi^0 & \frac{-2\Lambda}{\sqrt{6}}
    \end{pmatrix}
    \,.
\end{equation}
The resulting effective theory is valid for processes in which the momentum transfer is much smaller than the baryon mass.

Neglecting interactions with mesons, the leading terms in the effective Lagrangian that involve the sources $s_\chi$ and $s_G$ are
\begin{equation} \label{eq:L_HBChPT}
    \mathcal{L}_\text{HBChPT} \supset (b_0 \Tr[M_q] s_\chi - \frac{2}{27}m_G s_G ) \Tr(\bar{B}_v B_v) \,.
\end{equation}
We refer the reader to Ref.~\cite{Bishara:2016hek} for further details. The low-energy constants are related to nucleon matrix elements according to 
\begin{align}
    b_0 &= -\frac{1}{2} \langle p|\bar{u}{u}|p\rangle \,, \\
    m_G &= m_p - \sum_{q=u,d,s} m_q \langle p|\bar{q}{q}|p\rangle \,.
\end{align}
Using the most recent lattice determinations of the matrix elements from the FLAG review~\cite{FlavourLatticeAveragingGroupFLAG:2021npn,Harris:2019bih,BMW:2011sbi,Durr:2015dna,Yang:2015uis,Freeman:2012ry,Junnarkar:2013ac}, together with the expression for $\langle p|m_u\bar{u}{u}|p\rangle$ in Ref.~\cite{Durr:2015dna}, we obtain $m_G=(847\pm8)\,$MeV and $b_0=-3.2\pm0.3$. 

Using the Lagrangian \eqref{eq:L_HBChPT} together with the expressions for the sources in \cref{eq:sources_G,eq:sources_q}, the dark-matter--nucleon cross-section can be straightforwardly computed. The result is given by \cref{eq:sigma_chiN}.


\section{\texorpdfstring{$K\to\pi\chi\chi$}{K-meson decay}}
\label{app:meson-decay}

In this appendix, we provide further details of the calculation of the $K^+ \to \pi^+ \bar\chi \chi$ decay rate. The amplitudes describing the decay are closely related to those for $K$ decays into a light Higgs~\cite{Leutwyler:1989xj}.

Within the low-energy meson theory, the $s\to d$ transition is described by the $\Delta S=1$ effective weak Lagrangian. The leading contribution is from the term
\begin{align}
    \mathcal{L}^\text{LO}_{\Delta S=1} &\supset \frac{G_F}{\sqrt{2}} V_{ud} V_{us}^* \,g_8 f^4 \Tr[\lambda (U^\dagger D^\mu U)(U^\dagger D_\mu U)] + \text{h.c.} \,, \\
    &\supset -\sqrt{2} G_F V_{ud} V_{us}^* \,g_8 f^2 (\partial^\mu\pi^-)(\partial_\mu K^+) + \text{h.c.} \,,\label{eq:weakLag}
\end{align}
where $\lambda=(\lambda_6-i\lambda_7)/2$, with $\lambda_{6,7}$ Gell-Mann matrices, $G_F$ is the Fermi constant, $V_{ij}$ are CKM matrix elements, and the low-energy constant $g_8\approx5.0$~\cite{Cirigliano:2011ny}.

\begin{figure}
    \centering
    \includegraphics[width=0.47\columnwidth]{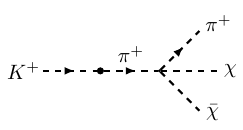}
    \hspace{1em}
    \includegraphics[width=0.47\columnwidth]{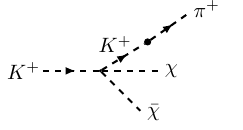}
    \caption{Feynman diagrams contributing to $K^+ \to \pi^+ \bar\chi \chi$ at leading order in the low-energy effective theory. \label{fig:Kdecay-IR}}
\end{figure}

Combining the Lagrangians in \cref{eq:weakLag,eq:L_ChPT2}, the leading contribution to the decay $K^+ \to \pi^+ \bar\chi \chi$ is given by the diagrams in \cref{fig:Kdecay-IR}. Note that the contribution to the dark matter vertex from the second line of \cref{eq:L_ChPT2} cancels between the two diagrams. Hence, at leading order, the decay rate is independent of the dark matter coupling to the light quarks.\footnote{This is the case only if the dark matter couples to \emph{all} three light quarks proportional to their masses.} For complex scalar dark matter, the resulting matrix element is
\begin{equation} \label{eq:Kpi_ME_IR_app}
   i\mathcal{M} = i \sqrt{2} G_F V_{ud} V_{us}^* g_8 f^2 \frac{c_G}{9\Lambda^2} (m_K^2 + m_\pi^2 - q^2) \,,
\end{equation}
where $q^2$ is the invariant mass of the pair of dark matter particles and $c_G=1$ or $c_G=-2$ in the gluon-coupled or quark-coupled cases, respectively.

\subsection{UV contribution in the quark-coupled case}

In quark-coupled models there is another contribution that arises in the UV. Here, we make the additional assumption that in the UV the operator $\mathcal{O}_\text{SM}^q$ includes couplings to all quarks proportional to their masses,
\begin{equation}
    \mathcal{O}_\text{SM}^q = \sum_q m_q \bar{q}{q} \,.
\end{equation}
Integrating out the top quark at one-loop then generates additional terms in the low-energy effective Lagrangian,
\begin{equation}
    \mathcal{L}_{sd} = c_{sd} \left( m_d \bar{s}_L d_R + m_s \bar{s}_R d_L \right) \mathcal{O}_\chi + \text{h.c.} \,,
\end{equation}
with\footnote{This expression differs from the well-known result for $K\to\pi h$, since here there is no coupling to the $W$ boson. This has previously been overlooked in the literature for models which couple only to the top quark~\cite{Knapen:2017xzo}.}
\begin{equation}
    c_{sd} = \frac{\sqrt{2} G_F m_t^2 V_{td} V_{ts}^*}{16\pi^2} F_t(m_W^2/m_t^2) \,,
\end{equation}
and
\begin{equation}
    F_t(x) = \frac{1 - 3x - 9x^2 + 11x^3 - 6x^2(1+x)\log x}{2(1-x)^3} \,.
\end{equation}

Following Ref.~\cite{Leutwyler:1989xj}, this matches onto the low-energy meson theory as
\begin{equation}
    \mathcal{L}_{sd} = -\frac{B_0 f^2}{2} s_\chi \mathrm{Tr} [\gamma M_q U + \text{h.c.}] \,,
\end{equation}
with
\begin{equation}
    \gamma = \begin{pmatrix}
    0 & 0 & 0 \\
    0 & 0 & c_{sd}^* \\
    0 & c_{sd} & 0
    \end{pmatrix} \,.
\end{equation}
The relevant terms are
\begin{equation}
    \mathcal{L}_{sd} \supset \frac{m_K^2}{2} s_\chi (c_{sd}\, \pi^- K^+ + \mathrm{h.c.} )\,,
\end{equation}
which, for scalar dark matter, give the following contribution to the decay matrix element
\begin{equation}
    i\mathcal{M}_{UV} = -i c_{sd} \frac{1}{2\Lambda^2} m_K^2 \,.
\end{equation}
One can easily check that this dominates over the IR contribution in \cref{eq:Kpi_ME_IR_app}.


\bibliography{DM-meson-BBN}

\begin{thebibliography}{67}%
\makeatletter
\providecommand \@ifxundefined [1]{%
 \@ifx{#1\undefined}
}%
\providecommand \@ifnum [1]{%
 \ifnum #1\expandafter \@firstoftwo
 \else \expandafter \@secondoftwo
 \fi
}%
\providecommand \@ifx [1]{%
 \ifx #1\expandafter \@firstoftwo
 \else \expandafter \@secondoftwo
 \fi
}%
\providecommand \natexlab [1]{#1}%
\providecommand \enquote  [1]{``#1''}%
\providecommand \bibnamefont  [1]{#1}%
\providecommand \bibfnamefont [1]{#1}%
\providecommand \citenamefont [1]{#1}%
\providecommand \href@noop [0]{\@secondoftwo}%
\providecommand \href [0]{\begingroup \@sanitize@url \@href}%
\providecommand \@href[1]{\@@startlink{#1}\@@href}%
\providecommand \@@href[1]{\endgroup#1\@@endlink}%
\providecommand \@sanitize@url [0]{\catcode `\\12\catcode `\$12\catcode
  `\&12\catcode `\#12\catcode `\^12\catcode `\_12\catcode `\%12\relax}%
\providecommand \@@startlink[1]{}%
\providecommand \@@endlink[0]{}%
\providecommand \url  [0]{\begingroup\@sanitize@url \@url }%
\providecommand \@url [1]{\endgroup\@href {#1}{\urlprefix }}%
\providecommand \urlprefix  [0]{URL }%
\providecommand \Eprint [0]{\href }%
\providecommand \doibase [0]{https://doi.org/}%
\providecommand \selectlanguage [0]{\@gobble}%
\providecommand \bibinfo  [0]{\@secondoftwo}%
\providecommand \bibfield  [0]{\@secondoftwo}%
\providecommand \translation [1]{[#1]}%
\providecommand \BibitemOpen [0]{}%
\providecommand \bibitemStop [0]{}%
\providecommand \bibitemNoStop [0]{.\EOS\space}%
\providecommand \EOS [0]{\spacefactor3000\relax}%
\providecommand \BibitemShut  [1]{\csname bibitem#1\endcsname}%
\let\auto@bib@innerbib\@empty
\bibitem [{\citenamefont {Kolb}\ \emph {et~al.}(1986)\citenamefont {Kolb},
  \citenamefont {Turner},\ and\ \citenamefont {Walker}}]{Kolb:1986nf}%
  \BibitemOpen
  \bibfield  {author} {\bibinfo {author} {\bibfnamefont {E.~W.}\ \bibnamefont
  {Kolb}}, \bibinfo {author} {\bibfnamefont {M.~S.}\ \bibnamefont {Turner}},\
  and\ \bibinfo {author} {\bibfnamefont {T.~P.}\ \bibnamefont {Walker}},\
  }\bibfield  {title} {\bibinfo {title} {{The Effect of Interacting Particles
  on Primordial Nucleosynthesis}},\ }\href
  {https://doi.org/10.1103/PhysRevD.34.2197} {\bibfield  {journal} {\bibinfo
  {journal} {Phys. Rev. D}\ }\textbf {\bibinfo {volume} {34}},\ \bibinfo
  {pages} {2197} (\bibinfo {year} {1986})}\BibitemShut {NoStop}%
\bibitem [{\citenamefont {Serpico}\ and\ \citenamefont
  {Raffelt}(2004)}]{Serpico:2004nm}%
  \BibitemOpen
  \bibfield  {author} {\bibinfo {author} {\bibfnamefont {P.~D.}\ \bibnamefont
  {Serpico}}\ and\ \bibinfo {author} {\bibfnamefont {G.~G.}\ \bibnamefont
  {Raffelt}},\ }\bibfield  {title} {\bibinfo {title} {{MeV-mass dark matter and
  primordial nucleosynthesis}},\ }\href
  {https://doi.org/10.1103/PhysRevD.70.043526} {\bibfield  {journal} {\bibinfo
  {journal} {Phys. Rev. D}\ }\textbf {\bibinfo {volume} {70}},\ \bibinfo
  {pages} {043526} (\bibinfo {year} {2004})},\ \Eprint
  {https://arxiv.org/abs/astro-ph/0403417} {arXiv:astro-ph/0403417}
  \BibitemShut {NoStop}%
\bibitem [{\citenamefont {Boehm}\ \emph {et~al.}(2012)\citenamefont {Boehm},
  \citenamefont {Dolan},\ and\ \citenamefont {McCabe}}]{Boehm:2012gr}%
  \BibitemOpen
  \bibfield  {author} {\bibinfo {author} {\bibfnamefont {C.}~\bibnamefont
  {Boehm}}, \bibinfo {author} {\bibfnamefont {M.~J.}\ \bibnamefont {Dolan}},\
  and\ \bibinfo {author} {\bibfnamefont {C.}~\bibnamefont {McCabe}},\
  }\bibfield  {title} {\bibinfo {title} {{Increasing Neff with particles in
  thermal equilibrium with neutrinos}},\ }\href
  {https://doi.org/10.1088/1475-7516/2012/12/027} {\bibfield  {journal}
  {\bibinfo  {journal} {JCAP}\ }\textbf {\bibinfo {volume} {12}}\bibfield
  {number} {\bibinfo  {number} { (2012)},\ \bibinfo {pages} {027}},\ }\Eprint
  {https://arxiv.org/abs/1207.0497} {arXiv:1207.0497 [astro-ph.CO]}
  \BibitemShut {NoStop}%
\bibitem [{\citenamefont {Berezhiani}\ \emph {et~al.}(2013)\citenamefont
  {Berezhiani}, \citenamefont {Dolgov},\ and\ \citenamefont
  {Tkachev}}]{Berezhiani:2012ru}%
  \BibitemOpen
  \bibfield  {author} {\bibinfo {author} {\bibfnamefont {Z.}~\bibnamefont
  {Berezhiani}}, \bibinfo {author} {\bibfnamefont {A.}~\bibnamefont {Dolgov}},\
  and\ \bibinfo {author} {\bibfnamefont {I.}~\bibnamefont {Tkachev}},\
  }\bibfield  {title} {\bibinfo {title} {{BBN with light dark matter}},\ }\href
  {https://doi.org/10.1088/1475-7516/2013/02/010} {\bibfield  {journal}
  {\bibinfo  {journal} {JCAP}\ }\textbf {\bibinfo {volume} {02}}\bibfield
  {number} {\bibinfo  {number} { (2013)},\ \bibinfo {pages} {010}},\ }\Eprint
  {https://arxiv.org/abs/1211.4937} {arXiv:1211.4937 [astro-ph.CO]}
  \BibitemShut {NoStop}%
\bibitem [{\citenamefont {Boehm}\ \emph {et~al.}(2013)\citenamefont {Boehm},
  \citenamefont {Dolan},\ and\ \citenamefont {McCabe}}]{Boehm:2013jpa}%
  \BibitemOpen
  \bibfield  {author} {\bibinfo {author} {\bibfnamefont {C.}~\bibnamefont
  {Boehm}}, \bibinfo {author} {\bibfnamefont {M.~J.}\ \bibnamefont {Dolan}},\
  and\ \bibinfo {author} {\bibfnamefont {C.}~\bibnamefont {McCabe}},\
  }\bibfield  {title} {\bibinfo {title} {{A Lower Bound on the Mass of Cold
  Thermal Dark Matter from Planck}},\ }\href
  {https://doi.org/10.1088/1475-7516/2013/08/041} {\bibfield  {journal}
  {\bibinfo  {journal} {JCAP}\ }\textbf {\bibinfo {volume} {08}}\bibfield
  {number} {\bibinfo  {number} { (2013)},\ \bibinfo {pages} {041}},\ }\Eprint
  {https://arxiv.org/abs/1303.6270} {arXiv:1303.6270 [hep-ph]} \BibitemShut
  {NoStop}%
\bibitem [{\citenamefont {Nollett}\ and\ \citenamefont
  {Steigman}(2014)}]{Nollett:2013pwa}%
  \BibitemOpen
  \bibfield  {author} {\bibinfo {author} {\bibfnamefont {K.~M.}\ \bibnamefont
  {Nollett}}\ and\ \bibinfo {author} {\bibfnamefont {G.}~\bibnamefont
  {Steigman}},\ }\bibfield  {title} {\bibinfo {title} {{BBN And The CMB
  Constrain Light, Electromagnetically Coupled WIMPs}},\ }\href
  {https://doi.org/10.1103/PhysRevD.89.083508} {\bibfield  {journal} {\bibinfo
  {journal} {Phys. Rev. D}\ }\textbf {\bibinfo {volume} {89}},\ \bibinfo
  {pages} {083508} (\bibinfo {year} {2014})},\ \Eprint
  {https://arxiv.org/abs/1312.5725} {arXiv:1312.5725 [astro-ph.CO]}
  \BibitemShut {NoStop}%
\bibitem [{\citenamefont {Steigman}\ and\ \citenamefont
  {Nollett}(2015)}]{Steigman:2014uqa}%
  \BibitemOpen
  \bibfield  {author} {\bibinfo {author} {\bibfnamefont {G.}~\bibnamefont
  {Steigman}}\ and\ \bibinfo {author} {\bibfnamefont {K.~M.}\ \bibnamefont
  {Nollett}},\ }\bibfield  {title} {\bibinfo {title} {{Light WIMPs And
  Equivalent Neutrinos}},\ }\href {https://doi.org/10.1016/j.phpro.2014.12.029}
  {\bibfield  {journal} {\bibinfo  {journal} {Phys. Procedia}\ }\textbf
  {\bibinfo {volume} {61}},\ \bibinfo {pages} {179} (\bibinfo {year} {2015})},\
  \Eprint {https://arxiv.org/abs/1402.5399} {arXiv:1402.5399 [astro-ph.CO]}
  \BibitemShut {NoStop}%
\bibitem [{\citenamefont {Nollett}\ and\ \citenamefont
  {Steigman}(2015)}]{Nollett:2014lwa}%
  \BibitemOpen
  \bibfield  {author} {\bibinfo {author} {\bibfnamefont {K.~M.}\ \bibnamefont
  {Nollett}}\ and\ \bibinfo {author} {\bibfnamefont {G.}~\bibnamefont
  {Steigman}},\ }\bibfield  {title} {\bibinfo {title} {{BBN And The CMB
  Constrain Neutrino Coupled Light WIMPs}},\ }\href
  {https://doi.org/10.1103/PhysRevD.91.083505} {\bibfield  {journal} {\bibinfo
  {journal} {Phys. Rev. D}\ }\textbf {\bibinfo {volume} {91}},\ \bibinfo
  {pages} {083505} (\bibinfo {year} {2015})},\ \Eprint
  {https://arxiv.org/abs/1411.6005} {arXiv:1411.6005 [astro-ph.CO]}
  \BibitemShut {NoStop}%
\bibitem [{\citenamefont {Kawasaki}\ \emph {et~al.}(2015)\citenamefont
  {Kawasaki}, \citenamefont {Kohri}, \citenamefont {Moroi},\ and\ \citenamefont
  {Takaesu}}]{Kawasaki:2015yya}%
  \BibitemOpen
  \bibfield  {author} {\bibinfo {author} {\bibfnamefont {M.}~\bibnamefont
  {Kawasaki}}, \bibinfo {author} {\bibfnamefont {K.}~\bibnamefont {Kohri}},
  \bibinfo {author} {\bibfnamefont {T.}~\bibnamefont {Moroi}},\ and\ \bibinfo
  {author} {\bibfnamefont {Y.}~\bibnamefont {Takaesu}},\ }\bibfield  {title}
  {\bibinfo {title} {{Revisiting Big-Bang Nucleosynthesis Constraints on
  Dark-Matter Annihilation}},\ }\href
  {https://doi.org/10.1016/j.physletb.2015.10.048} {\bibfield  {journal}
  {\bibinfo  {journal} {Phys. Lett. B}\ }\textbf {\bibinfo {volume} {751}},\
  \bibinfo {pages} {246} (\bibinfo {year} {2015})},\ \Eprint
  {https://arxiv.org/abs/1509.03665} {arXiv:1509.03665 [hep-ph]} \BibitemShut
  {NoStop}%
\bibitem [{\citenamefont {Wilkinson}\ \emph {et~al.}(2016)\citenamefont
  {Wilkinson}, \citenamefont {Vincent}, \citenamefont {B\oe{}hm},\ and\
  \citenamefont {McCabe}}]{Wilkinson:2016gsy}%
  \BibitemOpen
  \bibfield  {author} {\bibinfo {author} {\bibfnamefont {R.~J.}\ \bibnamefont
  {Wilkinson}}, \bibinfo {author} {\bibfnamefont {A.~C.}\ \bibnamefont
  {Vincent}}, \bibinfo {author} {\bibfnamefont {C.}~\bibnamefont {B\oe{}hm}},\
  and\ \bibinfo {author} {\bibfnamefont {C.}~\bibnamefont {McCabe}},\
  }\bibfield  {title} {\bibinfo {title} {{Ruling out the light weakly
  interacting massive particle explanation of the Galactic 511 keV line}},\
  }\href {https://doi.org/10.1103/PhysRevD.94.103525} {\bibfield  {journal}
  {\bibinfo  {journal} {Phys. Rev. D}\ }\textbf {\bibinfo {volume} {94}},\
  \bibinfo {pages} {103525} (\bibinfo {year} {2016})},\ \Eprint
  {https://arxiv.org/abs/1602.01114} {arXiv:1602.01114 [astro-ph.CO]}
  \BibitemShut {NoStop}%
\bibitem [{\citenamefont {Escudero}(2019)}]{Escudero:2018mvt}%
  \BibitemOpen
  \bibfield  {author} {\bibinfo {author} {\bibfnamefont {M.}~\bibnamefont
  {Escudero}},\ }\bibfield  {title} {\bibinfo {title} {{Neutrino decoupling
  beyond the Standard Model: CMB constraints on the Dark Matter mass with a
  fast and precise $N_{\rm eff}$ evaluation}},\ }\href
  {https://doi.org/10.1088/1475-7516/2019/02/007} {\bibfield  {journal}
  {\bibinfo  {journal} {JCAP}\ }\textbf {\bibinfo {volume} {02}}\bibfield
  {number} {\bibinfo  {number} { (2019)},\ \bibinfo {pages} {007}},\ }\Eprint
  {https://arxiv.org/abs/1812.05605} {arXiv:1812.05605 [hep-ph]} \BibitemShut
  {NoStop}%
\bibitem [{\citenamefont {Depta}\ \emph {et~al.}(2019)\citenamefont {Depta},
  \citenamefont {Hufnagel}, \citenamefont {Schmidt-Hoberg},\ and\ \citenamefont
  {Wild}}]{Depta:2019lbe}%
  \BibitemOpen
  \bibfield  {author} {\bibinfo {author} {\bibfnamefont {P.~F.}\ \bibnamefont
  {Depta}}, \bibinfo {author} {\bibfnamefont {M.}~\bibnamefont {Hufnagel}},
  \bibinfo {author} {\bibfnamefont {K.}~\bibnamefont {Schmidt-Hoberg}},\ and\
  \bibinfo {author} {\bibfnamefont {S.}~\bibnamefont {Wild}},\ }\bibfield
  {title} {\bibinfo {title} {{BBN constraints on the annihilation of MeV-scale
  dark matter}},\ }\href {https://doi.org/10.1088/1475-7516/2019/04/029}
  {\bibfield  {journal} {\bibinfo  {journal} {JCAP}\ }\textbf {\bibinfo
  {volume} {04}}\bibfield  {number} {\bibinfo  {number} { (2019)},\ \bibinfo
  {pages} {029}},\ }\Eprint {https://arxiv.org/abs/1901.06944}
  {arXiv:1901.06944 [hep-ph]} \BibitemShut {NoStop}%
\bibitem [{\citenamefont {Berlin}\ \emph {et~al.}(2019)\citenamefont {Berlin},
  \citenamefont {Blinov},\ and\ \citenamefont {Li}}]{Berlin:2019pbq}%
  \BibitemOpen
  \bibfield  {author} {\bibinfo {author} {\bibfnamefont {A.}~\bibnamefont
  {Berlin}}, \bibinfo {author} {\bibfnamefont {N.}~\bibnamefont {Blinov}},\
  and\ \bibinfo {author} {\bibfnamefont {S.~W.}\ \bibnamefont {Li}},\
  }\bibfield  {title} {\bibinfo {title} {{Dark Sector Equilibration During
  Nucleosynthesis}},\ }\href {https://doi.org/10.1103/PhysRevD.100.015038}
  {\bibfield  {journal} {\bibinfo  {journal} {Phys. Rev. D}\ }\textbf {\bibinfo
  {volume} {100}},\ \bibinfo {pages} {015038} (\bibinfo {year} {2019})},\
  \Eprint {https://arxiv.org/abs/1904.04256} {arXiv:1904.04256 [hep-ph]}
  \BibitemShut {NoStop}%
\bibitem [{\citenamefont {Sabti}\ \emph {et~al.}(2020)\citenamefont {Sabti},
  \citenamefont {Alvey}, \citenamefont {Escudero}, \citenamefont {Fairbairn},\
  and\ \citenamefont {Blas}}]{Sabti:2019mhn}%
  \BibitemOpen
  \bibfield  {author} {\bibinfo {author} {\bibfnamefont {N.}~\bibnamefont
  {Sabti}}, \bibinfo {author} {\bibfnamefont {J.}~\bibnamefont {Alvey}},
  \bibinfo {author} {\bibfnamefont {M.}~\bibnamefont {Escudero}}, \bibinfo
  {author} {\bibfnamefont {M.}~\bibnamefont {Fairbairn}},\ and\ \bibinfo
  {author} {\bibfnamefont {D.}~\bibnamefont {Blas}},\ }\bibfield  {title}
  {\bibinfo {title} {{Refined Bounds on MeV-scale Thermal Dark Sectors from BBN
  and the CMB}},\ }\href {https://doi.org/10.1088/1475-7516/2020/01/004}
  {\bibfield  {journal} {\bibinfo  {journal} {JCAP}\ }\textbf {\bibinfo
  {volume} {01}}\bibfield  {number} {\bibinfo  {number} { (2020)},\ \bibinfo
  {pages} {004}},\ }\Eprint {https://arxiv.org/abs/1910.01649}
  {arXiv:1910.01649 [hep-ph]} \BibitemShut {NoStop}%
\bibitem [{\citenamefont {Sabti}\ \emph {et~al.}(2021)\citenamefont {Sabti},
  \citenamefont {Alvey}, \citenamefont {Escudero}, \citenamefont {Fairbairn},\
  and\ \citenamefont {Blas}}]{Sabti:2021reh}%
  \BibitemOpen
  \bibfield  {author} {\bibinfo {author} {\bibfnamefont {N.}~\bibnamefont
  {Sabti}}, \bibinfo {author} {\bibfnamefont {J.}~\bibnamefont {Alvey}},
  \bibinfo {author} {\bibfnamefont {M.}~\bibnamefont {Escudero}}, \bibinfo
  {author} {\bibfnamefont {M.}~\bibnamefont {Fairbairn}},\ and\ \bibinfo
  {author} {\bibfnamefont {D.}~\bibnamefont {Blas}},\ }\bibfield  {title}
  {\bibinfo {title} {{Addendum: Refined bounds on MeV-scale thermal dark
  sectors from BBN and the CMB}},\ }\href
  {https://doi.org/10.1088/1475-7516/2021/08/A01} {\bibfield  {journal}
  {\bibinfo  {journal} {JCAP}\ }\textbf {\bibinfo {volume} {08}}\bibfield
  {number} {\bibinfo  {number} { (2021)},\ \bibinfo {pages} {A01}},\ }\Eprint
  {https://arxiv.org/abs/2107.11232} {arXiv:2107.11232 [hep-ph]} \BibitemShut
  {NoStop}%
\bibitem [{\citenamefont {Giovanetti}\ \emph {et~al.}(2022)\citenamefont
  {Giovanetti}, \citenamefont {Lisanti}, \citenamefont {Liu},\ and\
  \citenamefont {Ruderman}}]{Giovanetti:2021izc}%
  \BibitemOpen
  \bibfield  {author} {\bibinfo {author} {\bibfnamefont {C.}~\bibnamefont
  {Giovanetti}}, \bibinfo {author} {\bibfnamefont {M.}~\bibnamefont {Lisanti}},
  \bibinfo {author} {\bibfnamefont {H.}~\bibnamefont {Liu}},\ and\ \bibinfo
  {author} {\bibfnamefont {J.~T.}\ \bibnamefont {Ruderman}},\ }\bibfield
  {title} {\bibinfo {title} {{Joint Cosmic Microwave Background and Big Bang
  Nucleosynthesis Constraints on Light Dark Sectors with Dark Radiation}},\
  }\href {https://doi.org/10.1103/PhysRevLett.129.021302} {\bibfield  {journal}
  {\bibinfo  {journal} {Phys. Rev. Lett.}\ }\textbf {\bibinfo {volume} {129}},\
  \bibinfo {pages} {021302} (\bibinfo {year} {2022})},\ \Eprint
  {https://arxiv.org/abs/2109.03246} {arXiv:2109.03246 [hep-ph]} \BibitemShut
  {NoStop}%
\bibitem [{\citenamefont {Chu}\ \emph {et~al.}(2022)\citenamefont {Chu},
  \citenamefont {Kuo},\ and\ \citenamefont {Pradler}}]{Chu:2022xuh}%
  \BibitemOpen
  \bibfield  {author} {\bibinfo {author} {\bibfnamefont {X.}~\bibnamefont
  {Chu}}, \bibinfo {author} {\bibfnamefont {J.-L.}\ \bibnamefont {Kuo}},\ and\
  \bibinfo {author} {\bibfnamefont {J.}~\bibnamefont {Pradler}},\ }\bibfield
  {title} {\bibinfo {title} {{Toward a full description of MeV dark matter
  decoupling: A self-consistent determination of relic abundance and Neff}},\
  }\href {https://doi.org/10.1103/PhysRevD.106.055022} {\bibfield  {journal}
  {\bibinfo  {journal} {Phys. Rev. D}\ }\textbf {\bibinfo {volume} {106}},\
  \bibinfo {pages} {055022} (\bibinfo {year} {2022})},\ \Eprint
  {https://arxiv.org/abs/2205.05714} {arXiv:2205.05714 [hep-ph]} \BibitemShut
  {NoStop}%
\bibitem [{\citenamefont {An}\ \emph {et~al.}(2022)\citenamefont {An},
  \citenamefont {Gluscevic}, \citenamefont {Calabrese},\ and\ \citenamefont
  {Hill}}]{An:2022sva}%
  \BibitemOpen
  \bibfield  {author} {\bibinfo {author} {\bibfnamefont {R.}~\bibnamefont
  {An}}, \bibinfo {author} {\bibfnamefont {V.}~\bibnamefont {Gluscevic}},
  \bibinfo {author} {\bibfnamefont {E.}~\bibnamefont {Calabrese}},\ and\
  \bibinfo {author} {\bibfnamefont {J.~C.}\ \bibnamefont {Hill}},\ }\bibfield
  {title} {\bibinfo {title} {{What does cosmology tell us about the mass of
  thermal-relic dark matter?}},\ }\href
  {https://doi.org/10.1088/1475-7516/2022/07/002} {\bibfield  {journal}
  {\bibinfo  {journal} {JCAP}\ }\textbf {\bibinfo {volume} {07}}\bibfield
  {number} {\bibinfo  {number} { (2022)},\ \bibinfo {pages} {002}},\ }\Eprint
  {https://arxiv.org/abs/2202.03515} {arXiv:2202.03515 [astro-ph.CO]}
  \BibitemShut {NoStop}%
\bibitem [{\citenamefont {Knapen}\ \emph {et~al.}(2017)\citenamefont {Knapen},
  \citenamefont {Lin},\ and\ \citenamefont {Zurek}}]{Knapen:2017xzo}%
  \BibitemOpen
  \bibfield  {author} {\bibinfo {author} {\bibfnamefont {S.}~\bibnamefont
  {Knapen}}, \bibinfo {author} {\bibfnamefont {T.}~\bibnamefont {Lin}},\ and\
  \bibinfo {author} {\bibfnamefont {K.~M.}\ \bibnamefont {Zurek}},\ }\bibfield
  {title} {\bibinfo {title} {{Light Dark Matter: Models and Constraints}},\
  }\href {https://doi.org/10.1103/PhysRevD.96.115021} {\bibfield  {journal}
  {\bibinfo  {journal} {Phys. Rev. D}\ }\textbf {\bibinfo {volume} {96}},\
  \bibinfo {pages} {115021} (\bibinfo {year} {2017})},\ \Eprint
  {https://arxiv.org/abs/1709.07882} {arXiv:1709.07882 [hep-ph]} \BibitemShut
  {NoStop}%
\bibitem [{\citenamefont {Krnjaic}\ and\ \citenamefont
  {McDermott}(2020)}]{Krnjaic:2019dzc}%
  \BibitemOpen
  \bibfield  {author} {\bibinfo {author} {\bibfnamefont {G.}~\bibnamefont
  {Krnjaic}}\ and\ \bibinfo {author} {\bibfnamefont {S.~D.}\ \bibnamefont
  {McDermott}},\ }\bibfield  {title} {\bibinfo {title} {{Implications of BBN
  Bounds for Cosmic Ray Upscattered Dark Matter}},\ }\href
  {https://doi.org/10.1103/PhysRevD.101.123022} {\bibfield  {journal} {\bibinfo
   {journal} {Phys. Rev. D}\ }\textbf {\bibinfo {volume} {101}},\ \bibinfo
  {pages} {123022} (\bibinfo {year} {2020})},\ \Eprint
  {https://arxiv.org/abs/1908.00007} {arXiv:1908.00007 [hep-ph]} \BibitemShut
  {NoStop}%
\bibitem [{\citenamefont {Diamond}\ \emph {et~al.}(2024)\citenamefont
  {Diamond}, \citenamefont {Cappiello}, \citenamefont {Vincent},\ and\
  \citenamefont {Bramante}}]{Diamond:2023fsm}%
  \BibitemOpen
  \bibfield  {author} {\bibinfo {author} {\bibfnamefont {M.}~\bibnamefont
  {Diamond}}, \bibinfo {author} {\bibfnamefont {C.~V.}\ \bibnamefont
  {Cappiello}}, \bibinfo {author} {\bibfnamefont {A.~C.}\ \bibnamefont
  {Vincent}},\ and\ \bibinfo {author} {\bibfnamefont {J.}~\bibnamefont
  {Bramante}},\ }\bibfield  {title} {\bibinfo {title} {{Limiting Light Dark
  Matter with Luminous Hadronic Loops}},\ }\href
  {https://doi.org/10.1103/PhysRevLett.132.051001} {\bibfield  {journal}
  {\bibinfo  {journal} {Phys. Rev. Lett.}\ }\textbf {\bibinfo {volume} {132}},\
  \bibinfo {pages} {051001} (\bibinfo {year} {2024})},\ \Eprint
  {https://arxiv.org/abs/2307.13727} {arXiv:2307.13727 [hep-ph]} \BibitemShut
  {NoStop}%
\bibitem [{\citenamefont {Nadler}\ \emph {et~al.}(2019)\citenamefont {Nadler},
  \citenamefont {Gluscevic}, \citenamefont {Boddy},\ and\ \citenamefont
  {Wechsler}}]{Nadler:2019zrb}%
  \BibitemOpen
  \bibfield  {author} {\bibinfo {author} {\bibfnamefont {E.~O.}\ \bibnamefont
  {Nadler}}, \bibinfo {author} {\bibfnamefont {V.}~\bibnamefont {Gluscevic}},
  \bibinfo {author} {\bibfnamefont {K.~K.}\ \bibnamefont {Boddy}},\ and\
  \bibinfo {author} {\bibfnamefont {R.~H.}\ \bibnamefont {Wechsler}},\
  }\bibfield  {title} {\bibinfo {title} {{Constraints on Dark Matter
  Microphysics from the Milky Way Satellite Population}},\ }\href
  {https://doi.org/10.3847/2041-8213/ab1eb2} {\bibfield  {journal} {\bibinfo
  {journal} {Astrophys. J. Lett.}\ }\textbf {\bibinfo {volume} {878}},\
  \bibinfo {pages} {32} (\bibinfo {year} {2019})},\ \bibinfo {note} {[Erratum:
  Astrophys.J.Lett. 897, L46 (2020), Erratum: Astrophys.J. 897, L46 (2020)]},\
  \Eprint {https://arxiv.org/abs/1904.10000} {arXiv:1904.10000 [astro-ph.CO]}
  \BibitemShut {NoStop}%
\bibitem [{\citenamefont {Nadler}\ \emph {et~al.}(2021)\citenamefont {Nadler}
  \emph {et~al.}}]{DES:2020fxi}%
  \BibitemOpen
  \bibfield  {author} {\bibinfo {author} {\bibfnamefont {E.~O.}\ \bibnamefont
  {Nadler}} \emph {et~al.} (\bibinfo {collaboration} {DES}),\ }\bibfield
  {title} {\bibinfo {title} {{Milky Way Satellite Census. III. Constraints on
  Dark Matter Properties from Observations of Milky Way Satellite Galaxies}},\
  }\href {https://doi.org/10.1103/PhysRevLett.126.091101} {\bibfield  {journal}
  {\bibinfo  {journal} {Phys. Rev. Lett.}\ }\textbf {\bibinfo {volume} {126}},\
  \bibinfo {pages} {091101} (\bibinfo {year} {2021})},\ \Eprint
  {https://arxiv.org/abs/2008.00022} {arXiv:2008.00022 [astro-ph.CO]}
  \BibitemShut {NoStop}%
\bibitem [{\citenamefont {Maamari}\ \emph {et~al.}(2021)\citenamefont
  {Maamari}, \citenamefont {Gluscevic}, \citenamefont {Boddy}, \citenamefont
  {Nadler},\ and\ \citenamefont {Wechsler}}]{Maamari:2020aqz}%
  \BibitemOpen
  \bibfield  {author} {\bibinfo {author} {\bibfnamefont {K.}~\bibnamefont
  {Maamari}}, \bibinfo {author} {\bibfnamefont {V.}~\bibnamefont {Gluscevic}},
  \bibinfo {author} {\bibfnamefont {K.~K.}\ \bibnamefont {Boddy}}, \bibinfo
  {author} {\bibfnamefont {E.~O.}\ \bibnamefont {Nadler}},\ and\ \bibinfo
  {author} {\bibfnamefont {R.~H.}\ \bibnamefont {Wechsler}},\ }\bibfield
  {title} {\bibinfo {title} {{Bounds on velocity-dependent dark matter-proton
  scattering from Milky Way satellite abundance}},\ }\href
  {https://doi.org/10.3847/2041-8213/abd807} {\bibfield  {journal} {\bibinfo
  {journal} {Astrophys. J. Lett.}\ }\textbf {\bibinfo {volume} {907}},\
  \bibinfo {pages} {L46} (\bibinfo {year} {2021})},\ \Eprint
  {https://arxiv.org/abs/2010.02936} {arXiv:2010.02936 [astro-ph.CO]}
  \BibitemShut {NoStop}%
\bibitem [{\citenamefont {Buen-Abad}\ \emph {et~al.}(2022)\citenamefont
  {Buen-Abad}, \citenamefont {Essig}, \citenamefont {McKeen},\ and\
  \citenamefont {Zhong}}]{Buen-Abad:2021mvc}%
  \BibitemOpen
  \bibfield  {author} {\bibinfo {author} {\bibfnamefont {M.~A.}\ \bibnamefont
  {Buen-Abad}}, \bibinfo {author} {\bibfnamefont {R.}~\bibnamefont {Essig}},
  \bibinfo {author} {\bibfnamefont {D.}~\bibnamefont {McKeen}},\ and\ \bibinfo
  {author} {\bibfnamefont {Y.-M.}\ \bibnamefont {Zhong}},\ }\bibfield  {title}
  {\bibinfo {title} {{Cosmological constraints on dark matter interactions with
  ordinary matter}},\ }\href {https://doi.org/10.1016/j.physrep.2022.02.006}
  {\bibfield  {journal} {\bibinfo  {journal} {Phys. Rept.}\ }\textbf {\bibinfo
  {volume} {961}},\ \bibinfo {pages} {1} (\bibinfo {year} {2022})},\ \Eprint
  {https://arxiv.org/abs/2107.12377} {arXiv:2107.12377 [astro-ph.CO]}
  \BibitemShut {NoStop}%
\bibitem [{\citenamefont {Dvorkin}\ \emph {et~al.}(2014)\citenamefont
  {Dvorkin}, \citenamefont {Blum},\ and\ \citenamefont
  {Kamionkowski}}]{Dvorkin:2013cea}%
  \BibitemOpen
  \bibfield  {author} {\bibinfo {author} {\bibfnamefont {C.}~\bibnamefont
  {Dvorkin}}, \bibinfo {author} {\bibfnamefont {K.}~\bibnamefont {Blum}},\ and\
  \bibinfo {author} {\bibfnamefont {M.}~\bibnamefont {Kamionkowski}},\
  }\bibfield  {title} {\bibinfo {title} {{Constraining Dark Matter-Baryon
  Scattering with Linear Cosmology}},\ }\href
  {https://doi.org/10.1103/PhysRevD.89.023519} {\bibfield  {journal} {\bibinfo
  {journal} {Phys. Rev. D}\ }\textbf {\bibinfo {volume} {89}},\ \bibinfo
  {pages} {023519} (\bibinfo {year} {2014})},\ \Eprint
  {https://arxiv.org/abs/1311.2937} {arXiv:1311.2937 [astro-ph.CO]}
  \BibitemShut {NoStop}%
\bibitem [{\citenamefont {Xu}\ \emph {et~al.}(2018)\citenamefont {Xu},
  \citenamefont {Dvorkin},\ and\ \citenamefont {Chael}}]{Xu:2018efh}%
  \BibitemOpen
  \bibfield  {author} {\bibinfo {author} {\bibfnamefont {W.~L.}\ \bibnamefont
  {Xu}}, \bibinfo {author} {\bibfnamefont {C.}~\bibnamefont {Dvorkin}},\ and\
  \bibinfo {author} {\bibfnamefont {A.}~\bibnamefont {Chael}},\ }\bibfield
  {title} {\bibinfo {title} {{Probing sub-GeV Dark Matter-Baryon Scattering
  with Cosmological Observables}},\ }\href
  {https://doi.org/10.1103/PhysRevD.97.103530} {\bibfield  {journal} {\bibinfo
  {journal} {Phys. Rev. D}\ }\textbf {\bibinfo {volume} {97}},\ \bibinfo
  {pages} {103530} (\bibinfo {year} {2018})},\ \Eprint
  {https://arxiv.org/abs/1802.06788} {arXiv:1802.06788 [astro-ph.CO]}
  \BibitemShut {NoStop}%
\bibitem [{\citenamefont {Rogers}\ \emph {et~al.}(2022)\citenamefont {Rogers},
  \citenamefont {Dvorkin},\ and\ \citenamefont {Peiris}}]{Rogers:2021byl}%
  \BibitemOpen
  \bibfield  {author} {\bibinfo {author} {\bibfnamefont {K.~K.}\ \bibnamefont
  {Rogers}}, \bibinfo {author} {\bibfnamefont {C.}~\bibnamefont {Dvorkin}},\
  and\ \bibinfo {author} {\bibfnamefont {H.~V.}\ \bibnamefont {Peiris}},\
  }\bibfield  {title} {\bibinfo {title} {{Limits on the Light Dark
  Matter\textendash{}Proton Cross Section from Cosmic Large-Scale Structure}},\
  }\href {https://doi.org/10.1103/PhysRevLett.128.171301} {\bibfield  {journal}
  {\bibinfo  {journal} {Phys. Rev. Lett.}\ }\textbf {\bibinfo {volume} {128}},\
  \bibinfo {pages} {171301} (\bibinfo {year} {2022})},\ \Eprint
  {https://arxiv.org/abs/2111.10386} {arXiv:2111.10386 [astro-ph.CO]}
  \BibitemShut {NoStop}%
\bibitem [{\citenamefont {Chen}\ \emph {et~al.}(2002)\citenamefont {Chen},
  \citenamefont {Hannestad},\ and\ \citenamefont {Scherrer}}]{Chen:2002yh}%
  \BibitemOpen
  \bibfield  {author} {\bibinfo {author} {\bibfnamefont {X.-l.}\ \bibnamefont
  {Chen}}, \bibinfo {author} {\bibfnamefont {S.}~\bibnamefont {Hannestad}},\
  and\ \bibinfo {author} {\bibfnamefont {R.~J.}\ \bibnamefont {Scherrer}},\
  }\bibfield  {title} {\bibinfo {title} {{Cosmic microwave background and large
  scale structure limits on the interaction between dark matter and baryons}},\
  }\href {https://doi.org/10.1103/PhysRevD.65.123515} {\bibfield  {journal}
  {\bibinfo  {journal} {Phys. Rev. D}\ }\textbf {\bibinfo {volume} {65}},\
  \bibinfo {pages} {123515} (\bibinfo {year} {2002})},\ \Eprint
  {https://arxiv.org/abs/astro-ph/0202496} {arXiv:astro-ph/0202496}
  \BibitemShut {NoStop}%
\bibitem [{\citenamefont {Gluscevic}\ and\ \citenamefont
  {Boddy}(2018)}]{Gluscevic:2017ywp}%
  \BibitemOpen
  \bibfield  {author} {\bibinfo {author} {\bibfnamefont {V.}~\bibnamefont
  {Gluscevic}}\ and\ \bibinfo {author} {\bibfnamefont {K.~K.}\ \bibnamefont
  {Boddy}},\ }\bibfield  {title} {\bibinfo {title} {{Constraints on Scattering
  of keV\textendash{}TeV Dark Matter with Protons in the Early Universe}},\
  }\href {https://doi.org/10.1103/PhysRevLett.121.081301} {\bibfield  {journal}
  {\bibinfo  {journal} {Phys. Rev. Lett.}\ }\textbf {\bibinfo {volume} {121}},\
  \bibinfo {pages} {081301} (\bibinfo {year} {2018})},\ \Eprint
  {https://arxiv.org/abs/1712.07133} {arXiv:1712.07133 [astro-ph.CO]}
  \BibitemShut {NoStop}%
\bibitem [{\citenamefont {Slatyer}\ and\ \citenamefont
  {Wu}(2018)}]{Slatyer:2018aqg}%
  \BibitemOpen
  \bibfield  {author} {\bibinfo {author} {\bibfnamefont {T.~R.}\ \bibnamefont
  {Slatyer}}\ and\ \bibinfo {author} {\bibfnamefont {C.-L.}\ \bibnamefont
  {Wu}},\ }\bibfield  {title} {\bibinfo {title} {{Early-Universe constraints on
  dark matter-baryon scattering and their implications for a global 21 cm
  signal}},\ }\href {https://doi.org/10.1103/PhysRevD.98.023013} {\bibfield
  {journal} {\bibinfo  {journal} {Phys. Rev. D}\ }\textbf {\bibinfo {volume}
  {98}},\ \bibinfo {pages} {023013} (\bibinfo {year} {2018})},\ \Eprint
  {https://arxiv.org/abs/1803.09734} {arXiv:1803.09734 [astro-ph.CO]}
  \BibitemShut {NoStop}%
\bibitem [{\citenamefont {Krnjaic}(2016)}]{Krnjaic:2015mbs}%
  \BibitemOpen
  \bibfield  {author} {\bibinfo {author} {\bibfnamefont {G.}~\bibnamefont
  {Krnjaic}},\ }\bibfield  {title} {\bibinfo {title} {{Probing Light Thermal
  Dark-Matter With a Higgs Portal Mediator}},\ }\href
  {https://doi.org/10.1103/PhysRevD.94.073009} {\bibfield  {journal} {\bibinfo
  {journal} {Phys. Rev. D}\ }\textbf {\bibinfo {volume} {94}},\ \bibinfo
  {pages} {073009} (\bibinfo {year} {2016})},\ \Eprint
  {https://arxiv.org/abs/1512.04119} {arXiv:1512.04119 [hep-ph]} \BibitemShut
  {NoStop}%
\bibitem [{\citenamefont {Green}\ and\ \citenamefont
  {Rajendran}(2017)}]{Green:2017ybv}%
  \BibitemOpen
  \bibfield  {author} {\bibinfo {author} {\bibfnamefont {D.}~\bibnamefont
  {Green}}\ and\ \bibinfo {author} {\bibfnamefont {S.}~\bibnamefont
  {Rajendran}},\ }\bibfield  {title} {\bibinfo {title} {{The Cosmology of
  Sub-MeV Dark Matter}},\ }\href {https://doi.org/10.1007/JHEP10(2017)013}
  {\bibfield  {journal} {\bibinfo  {journal} {JHEP}\ }\textbf {\bibinfo
  {volume} {10}}\bibfield  {number} {\bibinfo  {number} { (2017)},\ \bibinfo
  {pages} {013}},\ }\Eprint {https://arxiv.org/abs/1701.08750}
  {arXiv:1701.08750 [hep-ph]} \BibitemShut {NoStop}%
\bibitem [{\citenamefont {Bishara}\ \emph {et~al.}(2017)\citenamefont
  {Bishara}, \citenamefont {Brod}, \citenamefont {Grinstein},\ and\
  \citenamefont {Zupan}}]{Bishara:2016hek}%
  \BibitemOpen
  \bibfield  {author} {\bibinfo {author} {\bibfnamefont {F.}~\bibnamefont
  {Bishara}}, \bibinfo {author} {\bibfnamefont {J.}~\bibnamefont {Brod}},
  \bibinfo {author} {\bibfnamefont {B.}~\bibnamefont {Grinstein}},\ and\
  \bibinfo {author} {\bibfnamefont {J.}~\bibnamefont {Zupan}},\ }\bibfield
  {title} {\bibinfo {title} {{Chiral Effective Theory of Dark Matter Direct
  Detection}},\ }\href {https://doi.org/10.1088/1475-7516/2017/02/009}
  {\bibfield  {journal} {\bibinfo  {journal} {JCAP}\ }\textbf {\bibinfo
  {volume} {02}}\bibfield  {number} {\bibinfo  {number} { (2017)},\ \bibinfo
  {pages} {009}},\ }\Eprint {https://arxiv.org/abs/1611.00368}
  {arXiv:1611.00368 [hep-ph]} \BibitemShut {NoStop}%
\bibitem [{\citenamefont {Aoki}\ \emph {et~al.}(2022)\citenamefont {Aoki} \emph
  {et~al.}}]{FlavourLatticeAveragingGroupFLAG:2021npn}%
  \BibitemOpen
  \bibfield  {author} {\bibinfo {author} {\bibfnamefont {Y.}~\bibnamefont
  {Aoki}} \emph {et~al.} (\bibinfo {collaboration} {Flavour Lattice Averaging
  Group (FLAG)}),\ }\bibfield  {title} {\bibinfo {title} {{FLAG Review 2021}},\
  }\href {https://doi.org/10.1140/epjc/s10052-022-10536-1} {\bibfield
  {journal} {\bibinfo  {journal} {Eur. Phys. J. C}\ }\textbf {\bibinfo {volume}
  {82}},\ \bibinfo {pages} {869} (\bibinfo {year} {2022})},\ \Eprint
  {https://arxiv.org/abs/2111.09849} {arXiv:2111.09849 [hep-lat]} \BibitemShut
  {NoStop}%
\bibitem [{\citenamefont {Leutwyler}\ and\ \citenamefont
  {Shifman}(1989)}]{Leutwyler:1989tn}%
  \BibitemOpen
  \bibfield  {author} {\bibinfo {author} {\bibfnamefont {H.}~\bibnamefont
  {Leutwyler}}\ and\ \bibinfo {author} {\bibfnamefont {M.~A.}\ \bibnamefont
  {Shifman}},\ }\bibfield  {title} {\bibinfo {title} {{Goldstone Bosons
  Generate Peculiar Conformal Anomalies}},\ }\href
  {https://doi.org/10.1016/0370-2693(89)91730-9} {\bibfield  {journal}
  {\bibinfo  {journal} {Phys. Lett. B}\ }\textbf {\bibinfo {volume} {221}},\
  \bibinfo {pages} {384} (\bibinfo {year} {1989})}\BibitemShut {NoStop}%
\bibitem [{\citenamefont {Workman}\ \emph {et~al.}(2022)\citenamefont {Workman}
  \emph {et~al.}}]{ParticleDataGroup:2022pth}%
  \BibitemOpen
  \bibfield  {author} {\bibinfo {author} {\bibfnamefont {R.~L.}\ \bibnamefont
  {Workman}} \emph {et~al.} (\bibinfo {collaboration} {Particle Data Group}),\
  }\bibfield  {title} {\bibinfo {title} {{Review of Particle Physics}},\ }\href
  {https://doi.org/10.1093/ptep/ptac097} {\bibfield  {journal} {\bibinfo
  {journal} {PTEP}\ }\textbf {\bibinfo {volume} {2022}},\ \bibinfo {pages}
  {083C01} (\bibinfo {year} {2022})}\BibitemShut {NoStop}%
\bibitem [{\citenamefont {Steigman}\ \emph {et~al.}(1977)\citenamefont
  {Steigman}, \citenamefont {Schramm},\ and\ \citenamefont
  {Gunn}}]{Steigman:1977kc}%
  \BibitemOpen
  \bibfield  {author} {\bibinfo {author} {\bibfnamefont {G.}~\bibnamefont
  {Steigman}}, \bibinfo {author} {\bibfnamefont {D.~N.}\ \bibnamefont
  {Schramm}},\ and\ \bibinfo {author} {\bibfnamefont {J.~E.}\ \bibnamefont
  {Gunn}},\ }\bibfield  {title} {\bibinfo {title} {{Cosmological Limits to the
  Number of Massive Leptons}},\ }\href
  {https://doi.org/10.1016/0370-2693(77)90176-9} {\bibfield  {journal}
  {\bibinfo  {journal} {Phys. Lett. B}\ }\textbf {\bibinfo {volume} {66}},\
  \bibinfo {pages} {202} (\bibinfo {year} {1977})}\BibitemShut {NoStop}%
\bibitem [{\citenamefont {Yeh}\ \emph {et~al.}(2022)\citenamefont {Yeh},
  \citenamefont {Shelton}, \citenamefont {Olive},\ and\ \citenamefont
  {Fields}}]{Yeh:2022heq}%
  \BibitemOpen
  \bibfield  {author} {\bibinfo {author} {\bibfnamefont {T.-H.}\ \bibnamefont
  {Yeh}}, \bibinfo {author} {\bibfnamefont {J.}~\bibnamefont {Shelton}},
  \bibinfo {author} {\bibfnamefont {K.~A.}\ \bibnamefont {Olive}},\ and\
  \bibinfo {author} {\bibfnamefont {B.~D.}\ \bibnamefont {Fields}},\ }\bibfield
   {title} {\bibinfo {title} {{Probing physics beyond the standard model:
  limits from BBN and the CMB independently and combined}},\ }\href
  {https://doi.org/10.1088/1475-7516/2022/10/046} {\bibfield  {journal}
  {\bibinfo  {journal} {JCAP}\ }\textbf {\bibinfo {volume} {10}}\bibfield
  {number} {\bibinfo  {number} { (2022)},\ \bibinfo {pages} {046}},\ }\Eprint
  {https://arxiv.org/abs/2207.13133} {arXiv:2207.13133 [astro-ph.CO]}
  \BibitemShut {NoStop}%
\bibitem [{\citenamefont {Aghanim}\ \emph {et~al.}(2020)\citenamefont {Aghanim}
  \emph {et~al.}}]{Planck:2018vyg}%
  \BibitemOpen
  \bibfield  {author} {\bibinfo {author} {\bibfnamefont {N.}~\bibnamefont
  {Aghanim}} \emph {et~al.} (\bibinfo {collaboration} {Planck}),\ }\bibfield
  {title} {\bibinfo {title} {{Planck 2018 results. VI. Cosmological
  parameters}},\ }\href {https://doi.org/10.1051/0004-6361/201833910}
  {\bibfield  {journal} {\bibinfo  {journal} {Astron. Astrophys.}\ }\textbf
  {\bibinfo {volume} {641}},\ \bibinfo {pages} {A6} (\bibinfo {year} {2020})},\
  \bibinfo {note} {[Erratum: Astron.Astrophys. 652, C4 (2021)]},\ \Eprint
  {https://arxiv.org/abs/1807.06209} {arXiv:1807.06209 [astro-ph.CO]}
  \BibitemShut {NoStop}%
\bibitem [{\citenamefont {Bird}\ \emph {et~al.}(2004)\citenamefont {Bird},
  \citenamefont {Jackson}, \citenamefont {Kowalewski},\ and\ \citenamefont
  {Pospelov}}]{Bird:2004ts}%
  \BibitemOpen
  \bibfield  {author} {\bibinfo {author} {\bibfnamefont {C.}~\bibnamefont
  {Bird}}, \bibinfo {author} {\bibfnamefont {P.}~\bibnamefont {Jackson}},
  \bibinfo {author} {\bibfnamefont {R.~V.}\ \bibnamefont {Kowalewski}},\ and\
  \bibinfo {author} {\bibfnamefont {M.}~\bibnamefont {Pospelov}},\ }\bibfield
  {title} {\bibinfo {title} {{Search for dark matter in b ---\ensuremath{>} s
  transitions with missing energy}},\ }\href
  {https://doi.org/10.1103/PhysRevLett.93.201803} {\bibfield  {journal}
  {\bibinfo  {journal} {Phys. Rev. Lett.}\ }\textbf {\bibinfo {volume} {93}},\
  \bibinfo {pages} {201803} (\bibinfo {year} {2004})},\ \Eprint
  {https://arxiv.org/abs/hep-ph/0401195} {arXiv:hep-ph/0401195} \BibitemShut
  {NoStop}%
\bibitem [{\citenamefont {Cortina~Gil}\ \emph {et~al.}(2021)\citenamefont
  {Cortina~Gil} \emph {et~al.}}]{NA62:2021zjw}%
  \BibitemOpen
  \bibfield  {author} {\bibinfo {author} {\bibfnamefont {E.}~\bibnamefont
  {Cortina~Gil}} \emph {et~al.} (\bibinfo {collaboration} {NA62}),\ }\bibfield
  {title} {\bibinfo {title} {{Measurement of the very rare
  K$^{+}$\textrightarrow{}$ {\pi}^{+}\nu \overline{\nu} $ decay}},\ }\href
  {https://doi.org/10.1007/JHEP06(2021)093} {\bibfield  {journal} {\bibinfo
  {journal} {JHEP}\ }\textbf {\bibinfo {volume} {06}}\bibfield  {number}
  {\bibinfo  {number} { (2021)},\ \bibinfo {pages} {093}},\ }\Eprint
  {https://arxiv.org/abs/2103.15389} {arXiv:2103.15389 [hep-ex]} \BibitemShut
  {NoStop}%
\bibitem [{\citenamefont {Pich}\ and\ \citenamefont
  {Rodr{\'\i}guez-S{\'a}nchez}(2021)}]{Pich:2021yll}%
  \BibitemOpen
  \bibfield  {author} {\bibinfo {author} {\bibfnamefont {A.}~\bibnamefont
  {Pich}}\ and\ \bibinfo {author} {\bibfnamefont {A.}~\bibnamefont
  {Rodr{\'\i}guez-S{\'a}nchez}},\ }\bibfield  {title} {\bibinfo {title} {{SU(3)
  analysis of four-quark operators: $K\to\pi\pi$ and vacuum matrix elements}},\
  }\href {https://doi.org/10.1007/JHEP06(2021)005} {\bibfield  {journal}
  {\bibinfo  {journal} {JHEP}\ }\textbf {\bibinfo {volume} {06}}\bibfield
  {number} {\bibinfo  {number} { (2021)},\ \bibinfo {pages} {005}},\ }\Eprint
  {https://arxiv.org/abs/2102.09308} {arXiv:2102.09308 [hep-ph]} \BibitemShut
  {NoStop}%
\bibitem [{\citenamefont {de~Salas}\ \emph {et~al.}(2015)\citenamefont
  {de~Salas}, \citenamefont {Lattanzi}, \citenamefont {Mangano}, \citenamefont
  {Miele}, \citenamefont {Pastor},\ and\ \citenamefont
  {Pisanti}}]{deSalas:2015glj}%
  \BibitemOpen
  \bibfield  {author} {\bibinfo {author} {\bibfnamefont {P.~F.}\ \bibnamefont
  {de~Salas}}, \bibinfo {author} {\bibfnamefont {M.}~\bibnamefont {Lattanzi}},
  \bibinfo {author} {\bibfnamefont {G.}~\bibnamefont {Mangano}}, \bibinfo
  {author} {\bibfnamefont {G.}~\bibnamefont {Miele}}, \bibinfo {author}
  {\bibfnamefont {S.}~\bibnamefont {Pastor}},\ and\ \bibinfo {author}
  {\bibfnamefont {O.}~\bibnamefont {Pisanti}},\ }\bibfield  {title} {\bibinfo
  {title} {{Bounds on very low reheating scenarios after Planck}},\ }\href
  {https://doi.org/10.1103/PhysRevD.92.123534} {\bibfield  {journal} {\bibinfo
  {journal} {Phys. Rev. D}\ }\textbf {\bibinfo {volume} {92}},\ \bibinfo
  {pages} {123534} (\bibinfo {year} {2015})},\ \Eprint
  {https://arxiv.org/abs/1511.00672} {arXiv:1511.00672 [astro-ph.CO]}
  \BibitemShut {NoStop}%
\bibitem [{\citenamefont {Hasegawa}\ \emph {et~al.}(2019)\citenamefont
  {Hasegawa}, \citenamefont {Hiroshima}, \citenamefont {Kohri}, \citenamefont
  {Hansen}, \citenamefont {Tram},\ and\ \citenamefont
  {Hannestad}}]{Hasegawa:2019jsa}%
  \BibitemOpen
  \bibfield  {author} {\bibinfo {author} {\bibfnamefont {T.}~\bibnamefont
  {Hasegawa}}, \bibinfo {author} {\bibfnamefont {N.}~\bibnamefont {Hiroshima}},
  \bibinfo {author} {\bibfnamefont {K.}~\bibnamefont {Kohri}}, \bibinfo
  {author} {\bibfnamefont {R.~S.~L.}\ \bibnamefont {Hansen}}, \bibinfo {author}
  {\bibfnamefont {T.}~\bibnamefont {Tram}},\ and\ \bibinfo {author}
  {\bibfnamefont {S.}~\bibnamefont {Hannestad}},\ }\bibfield  {title} {\bibinfo
  {title} {{MeV-scale reheating temperature and thermalization of oscillating
  neutrinos by radiative and hadronic decays of massive particles}},\ }\href
  {https://doi.org/10.1088/1475-7516/2019/12/012} {\bibfield  {journal}
  {\bibinfo  {journal} {JCAP}\ }\textbf {\bibinfo {volume} {12}}\bibfield
  {number} {\bibinfo  {number} { (2019)},\ \bibinfo {pages} {012}},\ }\Eprint
  {https://arxiv.org/abs/1908.10189} {arXiv:1908.10189 [hep-ph]} \BibitemShut
  {NoStop}%
\bibitem [{\citenamefont {Adari}\ \emph {et~al.}(2025)\citenamefont {Adari}
  \emph {et~al.}}]{SENSEI:2023zdf}%
  \BibitemOpen
  \bibfield  {author} {\bibinfo {author} {\bibfnamefont {P.}~\bibnamefont
  {Adari}} \emph {et~al.} (\bibinfo {collaboration} {SENSEI}),\ }\bibfield
  {title} {\bibinfo {title} {{First Direct-Detection Results on Sub-GeV Dark
  Matter Using the SENSEI Detector at SNOLAB}},\ }\href
  {https://doi.org/10.1103/PhysRevLett.134.011804} {\bibfield  {journal}
  {\bibinfo  {journal} {Phys. Rev. Lett.}\ }\textbf {\bibinfo {volume} {134}},\
  \bibinfo {pages} {011804} (\bibinfo {year} {2025})},\ \Eprint
  {https://arxiv.org/abs/2312.13342} {arXiv:2312.13342 [astro-ph.CO]}
  \BibitemShut {NoStop}%
\bibitem [{\citenamefont {Huang}\ \emph {et~al.}(2023)\citenamefont {Huang}
  \emph {et~al.}}]{PandaX:2023xgl}%
  \BibitemOpen
  \bibfield  {author} {\bibinfo {author} {\bibfnamefont {D.}~\bibnamefont
  {Huang}} \emph {et~al.} (\bibinfo {collaboration} {PandaX}),\ }\bibfield
  {title} {\bibinfo {title} {{Search for Dark-Matter\textendash{}Nucleon
  Interactions with a Dark Mediator in PandaX-4T}},\ }\href
  {https://doi.org/10.1103/PhysRevLett.131.191002} {\bibfield  {journal}
  {\bibinfo  {journal} {Phys. Rev. Lett.}\ }\textbf {\bibinfo {volume} {131}},\
  \bibinfo {pages} {191002} (\bibinfo {year} {2023})},\ \Eprint
  {https://arxiv.org/abs/2308.01540} {arXiv:2308.01540 [hep-ex]} \BibitemShut
  {NoStop}%
\bibitem [{\citenamefont {Bhattiprolu}\ \emph {et~al.}(2023)\citenamefont
  {Bhattiprolu}, \citenamefont {Elor}, \citenamefont {McGehee},\ and\
  \citenamefont {Pierce}}]{Bhattiprolu:2022sdd}%
  \BibitemOpen
  \bibfield  {author} {\bibinfo {author} {\bibfnamefont {P.~N.}\ \bibnamefont
  {Bhattiprolu}}, \bibinfo {author} {\bibfnamefont {G.}~\bibnamefont {Elor}},
  \bibinfo {author} {\bibfnamefont {R.}~\bibnamefont {McGehee}},\ and\ \bibinfo
  {author} {\bibfnamefont {A.}~\bibnamefont {Pierce}},\ }\bibfield  {title}
  {\bibinfo {title} {{Freezing-in hadrophilic dark matter at low reheating
  temperatures}},\ }\href {https://doi.org/10.1007/JHEP01(2023)128} {\bibfield
  {journal} {\bibinfo  {journal} {JHEP}\ }\textbf {\bibinfo {volume}
  {01}}\bibfield  {number} {\bibinfo  {number} { (2023)},\ \bibinfo {pages}
  {128}},\ }\Eprint {https://arxiv.org/abs/2210.15653} {arXiv:2210.15653
  [hep-ph]} \BibitemShut {NoStop}%
\bibitem [{\citenamefont {Ballesteros}\ \emph {et~al.}(2021)\citenamefont
  {Ballesteros}, \citenamefont {Garcia},\ and\ \citenamefont
  {Pierre}}]{Ballesteros:2020adh}%
  \BibitemOpen
  \bibfield  {author} {\bibinfo {author} {\bibfnamefont {G.}~\bibnamefont
  {Ballesteros}}, \bibinfo {author} {\bibfnamefont {M.~A.~G.}\ \bibnamefont
  {Garcia}},\ and\ \bibinfo {author} {\bibfnamefont {M.}~\bibnamefont
  {Pierre}},\ }\bibfield  {title} {\bibinfo {title} {{How warm are non-thermal
  relics? Lyman-$\alpha$ bounds on out-of-equilibrium dark matter}},\ }\href
  {https://doi.org/10.1088/1475-7516/2021/03/101} {\bibfield  {journal}
  {\bibinfo  {journal} {JCAP}\ }\textbf {\bibinfo {volume} {03}}\bibfield
  {number} {\bibinfo  {number} { (2021)},\ \bibinfo {pages} {101}},\ }\Eprint
  {https://arxiv.org/abs/2011.13458} {arXiv:2011.13458 [hep-ph]} \BibitemShut
  {NoStop}%
\bibitem [{\citenamefont {Hertel}\ \emph {et~al.}(2019)\citenamefont {Hertel},
  \citenamefont {Biekert}, \citenamefont {Lin}, \citenamefont {Velan},\ and\
  \citenamefont {McKinsey}}]{Hertel:2018aal}%
  \BibitemOpen
  \bibfield  {author} {\bibinfo {author} {\bibfnamefont {S.~A.}\ \bibnamefont
  {Hertel}}, \bibinfo {author} {\bibfnamefont {A.}~\bibnamefont {Biekert}},
  \bibinfo {author} {\bibfnamefont {J.}~\bibnamefont {Lin}}, \bibinfo {author}
  {\bibfnamefont {V.}~\bibnamefont {Velan}},\ and\ \bibinfo {author}
  {\bibfnamefont {D.~N.}\ \bibnamefont {McKinsey}},\ }\bibfield  {title}
  {\bibinfo {title} {{Direct detection of sub-GeV dark matter using a
  superfluid $^4$He target}},\ }\href
  {https://doi.org/10.1103/PhysRevD.100.092007} {\bibfield  {journal} {\bibinfo
   {journal} {Phys. Rev. D}\ }\textbf {\bibinfo {volume} {100}},\ \bibinfo
  {pages} {092007} (\bibinfo {year} {2019})},\ \Eprint
  {https://arxiv.org/abs/1810.06283} {arXiv:1810.06283 [physics.ins-det]}
  \BibitemShut {NoStop}%
\bibitem [{\citenamefont {Aguilar-Arevalo}\ \emph {et~al.}(2022)\citenamefont
  {Aguilar-Arevalo} \emph {et~al.}}]{Oscura:2022vmi}%
  \BibitemOpen
  \bibfield  {author} {\bibinfo {author} {\bibfnamefont {A.}~\bibnamefont
  {Aguilar-Arevalo}} \emph {et~al.} (\bibinfo {collaboration} {Oscura}),\
  }\bibfield  {title} {\bibinfo {title} {{The Oscura Experiment}},\ }\href@noop
  {} {\  (\bibinfo {year} {2022})},\ \Eprint {https://arxiv.org/abs/2202.10518}
  {arXiv:2202.10518 [astro-ph.IM]} \BibitemShut {NoStop}%
\bibitem [{\citenamefont {von Krosigk}\ \emph {et~al.}(2023)\citenamefont {von
  Krosigk} \emph {et~al.}}]{vonKrosigk:2022vnf}%
  \BibitemOpen
  \bibfield  {author} {\bibinfo {author} {\bibfnamefont {B.}~\bibnamefont {von
  Krosigk}} \emph {et~al.},\ }\bibfield  {title} {\bibinfo {title} {{DELight: A
  Direct search Experiment for Light dark matter with superfluid helium}},\
  }\href {https://doi.org/10.21468/SciPostPhysProc.12.016} {\bibfield
  {journal} {\bibinfo  {journal} {SciPost Phys. Proc.}\ }\textbf {\bibinfo
  {volume} {12}},\ \bibinfo {pages} {016} (\bibinfo {year} {2023})},\ \Eprint
  {https://arxiv.org/abs/2209.10950} {arXiv:2209.10950 [hep-ex]} \BibitemShut
  {NoStop}%
\bibitem [{\citenamefont {Balogh}\ \emph {et~al.}(2023)\citenamefont {Balogh}
  \emph {et~al.}}]{NEWS-G:2023qwh}%
  \BibitemOpen
  \bibfield  {author} {\bibinfo {author} {\bibfnamefont {L.}~\bibnamefont
  {Balogh}} \emph {et~al.} (\bibinfo {collaboration} {NEWS-G}),\ }\bibfield
  {title} {\bibinfo {title} {{Exploring light dark matter with the DarkSPHERE
  spherical proportional counter electroformed underground at the Boulby
  Underground Laboratory}},\ }\href
  {https://doi.org/10.1103/PhysRevD.108.112006} {\bibfield  {journal} {\bibinfo
   {journal} {Phys. Rev. D}\ }\textbf {\bibinfo {volume} {108}},\ \bibinfo
  {pages} {112006} (\bibinfo {year} {2023})},\ \Eprint
  {https://arxiv.org/abs/2301.05183} {arXiv:2301.05183 [hep-ex]} \BibitemShut
  {NoStop}%
\bibitem [{\citenamefont {Batell}\ \emph {et~al.}(2018)\citenamefont {Batell},
  \citenamefont {Freitas}, \citenamefont {Ismail},\ and\ \citenamefont
  {Mckeen}}]{Batell:2017kty}%
  \BibitemOpen
  \bibfield  {author} {\bibinfo {author} {\bibfnamefont {B.}~\bibnamefont
  {Batell}}, \bibinfo {author} {\bibfnamefont {A.}~\bibnamefont {Freitas}},
  \bibinfo {author} {\bibfnamefont {A.}~\bibnamefont {Ismail}},\ and\ \bibinfo
  {author} {\bibfnamefont {D.}~\bibnamefont {Mckeen}},\ }\bibfield  {title}
  {\bibinfo {title} {{Flavor-specific scalar mediators}},\ }\href
  {https://doi.org/10.1103/PhysRevD.98.055026} {\bibfield  {journal} {\bibinfo
  {journal} {Phys. Rev. D}\ }\textbf {\bibinfo {volume} {98}},\ \bibinfo
  {pages} {055026} (\bibinfo {year} {2018})},\ \Eprint
  {https://arxiv.org/abs/1712.10022} {arXiv:1712.10022 [hep-ph]} \BibitemShut
  {NoStop}%
\bibitem [{\citenamefont {Batell}\ \emph {et~al.}(2019)\citenamefont {Batell},
  \citenamefont {Freitas}, \citenamefont {Ismail},\ and\ \citenamefont
  {Mckeen}}]{Batell:2018fqo}%
  \BibitemOpen
  \bibfield  {author} {\bibinfo {author} {\bibfnamefont {B.}~\bibnamefont
  {Batell}}, \bibinfo {author} {\bibfnamefont {A.}~\bibnamefont {Freitas}},
  \bibinfo {author} {\bibfnamefont {A.}~\bibnamefont {Ismail}},\ and\ \bibinfo
  {author} {\bibfnamefont {D.}~\bibnamefont {Mckeen}},\ }\bibfield  {title}
  {\bibinfo {title} {{Probing Light Dark Matter with a Hadrophilic Scalar
  Mediator}},\ }\href {https://doi.org/10.1103/PhysRevD.100.095020} {\bibfield
  {journal} {\bibinfo  {journal} {Phys. Rev. D}\ }\textbf {\bibinfo {volume}
  {100}},\ \bibinfo {pages} {095020} (\bibinfo {year} {2019})},\ \Eprint
  {https://arxiv.org/abs/1812.05103} {arXiv:1812.05103 [hep-ph]} \BibitemShut
  {NoStop}%
\bibitem [{\citenamefont {Voloshin}\ and\ \citenamefont
  {Zakharov}(1980)}]{Voloshin:1980zf}%
  \BibitemOpen
  \bibfield  {author} {\bibinfo {author} {\bibfnamefont {M.~B.}\ \bibnamefont
  {Voloshin}}\ and\ \bibinfo {author} {\bibfnamefont {V.~I.}\ \bibnamefont
  {Zakharov}},\ }\bibfield  {title} {\bibinfo {title} {{Measuring QCD Anomalies
  in Hadronic Transitions Between Onium States}},\ }\href
  {https://doi.org/10.1103/PhysRevLett.45.688} {\bibfield  {journal} {\bibinfo
  {journal} {Phys. Rev. Lett.}\ }\textbf {\bibinfo {volume} {45}},\ \bibinfo
  {pages} {688} (\bibinfo {year} {1980})}\BibitemShut {NoStop}%
\bibitem [{\citenamefont {Chivukula}\ \emph {et~al.}(1989)\citenamefont
  {Chivukula}, \citenamefont {Cohen}, \citenamefont {Georgi}, \citenamefont
  {Grinstein},\ and\ \citenamefont {Manohar}}]{Chivukula:1989ds}%
  \BibitemOpen
  \bibfield  {author} {\bibinfo {author} {\bibfnamefont {R.~S.}\ \bibnamefont
  {Chivukula}}, \bibinfo {author} {\bibfnamefont {A.~G.}\ \bibnamefont
  {Cohen}}, \bibinfo {author} {\bibfnamefont {H.}~\bibnamefont {Georgi}},
  \bibinfo {author} {\bibfnamefont {B.}~\bibnamefont {Grinstein}},\ and\
  \bibinfo {author} {\bibfnamefont {A.~V.}\ \bibnamefont {Manohar}},\
  }\bibfield  {title} {\bibinfo {title} {{Higgs Decay Into Goldstone Bosons}},\
  }\href {https://doi.org/10.1016/0003-4916(89)90119-X} {\bibfield  {journal}
  {\bibinfo  {journal} {Annals Phys.}\ }\textbf {\bibinfo {volume} {192}},\
  \bibinfo {pages} {93} (\bibinfo {year} {1989})}\BibitemShut {NoStop}%
\bibitem [{\citenamefont {Donoghue}\ \emph {et~al.}(1990)\citenamefont
  {Donoghue}, \citenamefont {Gasser},\ and\ \citenamefont
  {Leutwyler}}]{Donoghue:1990xh}%
  \BibitemOpen
  \bibfield  {author} {\bibinfo {author} {\bibfnamefont {J.~F.}\ \bibnamefont
  {Donoghue}}, \bibinfo {author} {\bibfnamefont {J.}~\bibnamefont {Gasser}},\
  and\ \bibinfo {author} {\bibfnamefont {H.}~\bibnamefont {Leutwyler}},\
  }\bibfield  {title} {\bibinfo {title} {{The Decay of a Light Higgs Boson}},\
  }\href {https://doi.org/10.1016/0550-3213(90)90474-R} {\bibfield  {journal}
  {\bibinfo  {journal} {Nucl. Phys. B}\ }\textbf {\bibinfo {volume} {343}},\
  \bibinfo {pages} {341} (\bibinfo {year} {1990})}\BibitemShut {NoStop}%
\bibitem [{\citenamefont {Jenkins}\ and\ \citenamefont
  {Manohar}(1991)}]{Jenkins:1990jv}%
  \BibitemOpen
  \bibfield  {author} {\bibinfo {author} {\bibfnamefont {E.~E.}\ \bibnamefont
  {Jenkins}}\ and\ \bibinfo {author} {\bibfnamefont {A.~V.}\ \bibnamefont
  {Manohar}},\ }\bibfield  {title} {\bibinfo {title} {{Baryon chiral
  perturbation theory using a heavy fermion Lagrangian}},\ }\href
  {https://doi.org/10.1016/0370-2693(91)90266-S} {\bibfield  {journal}
  {\bibinfo  {journal} {Phys. Lett. B}\ }\textbf {\bibinfo {volume} {255}},\
  \bibinfo {pages} {558} (\bibinfo {year} {1991})}\BibitemShut {NoStop}%
\bibitem [{\citenamefont {Harris}\ \emph {et~al.}(2019)\citenamefont {Harris},
  \citenamefont {von Hippel}, \citenamefont {Junnarkar}, \citenamefont {Meyer},
  \citenamefont {Ottnad}, \citenamefont {Wilhelm}, \citenamefont {Wittig},\
  and\ \citenamefont {Wrang}}]{Harris:2019bih}%
  \BibitemOpen
  \bibfield  {author} {\bibinfo {author} {\bibfnamefont {T.}~\bibnamefont
  {Harris}}, \bibinfo {author} {\bibfnamefont {G.}~\bibnamefont {von Hippel}},
  \bibinfo {author} {\bibfnamefont {P.}~\bibnamefont {Junnarkar}}, \bibinfo
  {author} {\bibfnamefont {H.~B.}\ \bibnamefont {Meyer}}, \bibinfo {author}
  {\bibfnamefont {K.}~\bibnamefont {Ottnad}}, \bibinfo {author} {\bibfnamefont
  {J.}~\bibnamefont {Wilhelm}}, \bibinfo {author} {\bibfnamefont
  {H.}~\bibnamefont {Wittig}},\ and\ \bibinfo {author} {\bibfnamefont
  {L.}~\bibnamefont {Wrang}},\ }\bibfield  {title} {\bibinfo {title} {{Nucleon
  isovector charges and twist-2 matrix elements with $N_f$}},\ }\href
  {https://doi.org/10.1103/PhysRevD.100.034513} {\bibfield  {journal} {\bibinfo
   {journal} {Phys. Rev. D}\ }\textbf {\bibinfo {volume} {100}},\ \bibinfo
  {pages} {034513} (\bibinfo {year} {2019})},\ \Eprint
  {https://arxiv.org/abs/1905.01291} {arXiv:1905.01291 [hep-lat]} \BibitemShut
  {NoStop}%
\bibitem [{\citenamefont {Durr}\ \emph {et~al.}(2012)\citenamefont {Durr} \emph
  {et~al.}}]{BMW:2011sbi}%
  \BibitemOpen
  \bibfield  {author} {\bibinfo {author} {\bibfnamefont {S.}~\bibnamefont
  {Durr}} \emph {et~al.} (\bibinfo {collaboration} {BMW}),\ }\bibfield  {title}
  {\bibinfo {title} {{Sigma term and strangeness content of octet baryons}},\
  }\href {https://doi.org/10.1103/PhysRevD.85.014509} {\bibfield  {journal}
  {\bibinfo  {journal} {Phys. Rev. D}\ }\textbf {\bibinfo {volume} {85}},\
  \bibinfo {pages} {014509} (\bibinfo {year} {2012})},\ \bibinfo {note}
  {[Erratum: Phys.Rev.D 93, 039905 (2016)]},\ \Eprint
  {https://arxiv.org/abs/1109.4265} {arXiv:1109.4265 [hep-lat]} \BibitemShut
  {NoStop}%
\bibitem [{\citenamefont {Durr}\ \emph {et~al.}(2016)\citenamefont {Durr} \emph
  {et~al.}}]{Durr:2015dna}%
  \BibitemOpen
  \bibfield  {author} {\bibinfo {author} {\bibfnamefont {S.}~\bibnamefont
  {Durr}} \emph {et~al.},\ }\bibfield  {title} {\bibinfo {title} {{Lattice
  computation of the nucleon scalar quark contents at the physical point}},\
  }\href {https://doi.org/10.1103/PhysRevLett.116.172001} {\bibfield  {journal}
  {\bibinfo  {journal} {Phys. Rev. Lett.}\ }\textbf {\bibinfo {volume} {116}},\
  \bibinfo {pages} {172001} (\bibinfo {year} {2016})},\ \Eprint
  {https://arxiv.org/abs/1510.08013} {arXiv:1510.08013 [hep-lat]} \BibitemShut
  {NoStop}%
\bibitem [{\citenamefont {Yang}\ \emph {et~al.}(2016)\citenamefont {Yang},
  \citenamefont {Alexandru}, \citenamefont {Draper}, \citenamefont {Liang},\
  and\ \citenamefont {Liu}}]{Yang:2015uis}%
  \BibitemOpen
  \bibfield  {author} {\bibinfo {author} {\bibfnamefont {Y.-B.}\ \bibnamefont
  {Yang}}, \bibinfo {author} {\bibfnamefont {A.}~\bibnamefont {Alexandru}},
  \bibinfo {author} {\bibfnamefont {T.}~\bibnamefont {Draper}}, \bibinfo
  {author} {\bibfnamefont {J.}~\bibnamefont {Liang}},\ and\ \bibinfo {author}
  {\bibfnamefont {K.-F.}\ \bibnamefont {Liu}} (\bibinfo {collaboration}
  {xQCD}),\ }\bibfield  {title} {\bibinfo {title} {{$\pi$N and strangeness
  sigma terms at the physical point with chiral fermions}},\ }\href
  {https://doi.org/10.1103/PhysRevD.94.054503} {\bibfield  {journal} {\bibinfo
  {journal} {Phys. Rev. D}\ }\textbf {\bibinfo {volume} {94}},\ \bibinfo
  {pages} {054503} (\bibinfo {year} {2016})},\ \Eprint
  {https://arxiv.org/abs/1511.09089} {arXiv:1511.09089 [hep-lat]} \BibitemShut
  {NoStop}%
\bibitem [{\citenamefont {Freeman}\ and\ \citenamefont
  {Toussaint}(2013)}]{Freeman:2012ry}%
  \BibitemOpen
  \bibfield  {author} {\bibinfo {author} {\bibfnamefont {W.}~\bibnamefont
  {Freeman}}\ and\ \bibinfo {author} {\bibfnamefont {D.}~\bibnamefont
  {Toussaint}} (\bibinfo {collaboration} {MILC}),\ }\bibfield  {title}
  {\bibinfo {title} {{Intrinsic strangeness and charm of the nucleon using
  improved staggered fermions}},\ }\href
  {https://doi.org/10.1103/PhysRevD.88.054503} {\bibfield  {journal} {\bibinfo
  {journal} {Phys. Rev. D}\ }\textbf {\bibinfo {volume} {88}},\ \bibinfo
  {pages} {054503} (\bibinfo {year} {2013})},\ \Eprint
  {https://arxiv.org/abs/1204.3866} {arXiv:1204.3866 [hep-lat]} \BibitemShut
  {NoStop}%
\bibitem [{\citenamefont {Junnarkar}\ and\ \citenamefont
  {Walker-Loud}(2013)}]{Junnarkar:2013ac}%
  \BibitemOpen
  \bibfield  {author} {\bibinfo {author} {\bibfnamefont {P.}~\bibnamefont
  {Junnarkar}}\ and\ \bibinfo {author} {\bibfnamefont {A.}~\bibnamefont
  {Walker-Loud}},\ }\bibfield  {title} {\bibinfo {title} {{Scalar strange
  content of the nucleon from lattice QCD}},\ }\href
  {https://doi.org/10.1103/PhysRevD.87.114510} {\bibfield  {journal} {\bibinfo
  {journal} {Phys. Rev. D}\ }\textbf {\bibinfo {volume} {87}},\ \bibinfo
  {pages} {114510} (\bibinfo {year} {2013})},\ \Eprint
  {https://arxiv.org/abs/1301.1114} {arXiv:1301.1114 [hep-lat]} \BibitemShut
  {NoStop}%
\bibitem [{\citenamefont {Leutwyler}\ and\ \citenamefont
  {Shifman}(1990)}]{Leutwyler:1989xj}%
  \BibitemOpen
  \bibfield  {author} {\bibinfo {author} {\bibfnamefont {H.}~\bibnamefont
  {Leutwyler}}\ and\ \bibinfo {author} {\bibfnamefont {M.~A.}\ \bibnamefont
  {Shifman}},\ }\bibfield  {title} {\bibinfo {title} {{Light Higgs Particle in
  Decays of $K$ and $\eta$ Mesons}},\ }\href
  {https://doi.org/10.1016/0550-3213(90)90475-S} {\bibfield  {journal}
  {\bibinfo  {journal} {Nucl. Phys. B}\ }\textbf {\bibinfo {volume} {343}},\
  \bibinfo {pages} {369} (\bibinfo {year} {1990})}\BibitemShut {NoStop}%
\bibitem [{\citenamefont {Cirigliano}\ \emph {et~al.}(2012)\citenamefont
  {Cirigliano}, \citenamefont {Ecker}, \citenamefont {Neufeld}, \citenamefont
  {Pich},\ and\ \citenamefont {Portoles}}]{Cirigliano:2011ny}%
  \BibitemOpen
  \bibfield  {author} {\bibinfo {author} {\bibfnamefont {V.}~\bibnamefont
  {Cirigliano}}, \bibinfo {author} {\bibfnamefont {G.}~\bibnamefont {Ecker}},
  \bibinfo {author} {\bibfnamefont {H.}~\bibnamefont {Neufeld}}, \bibinfo
  {author} {\bibfnamefont {A.}~\bibnamefont {Pich}},\ and\ \bibinfo {author}
  {\bibfnamefont {J.}~\bibnamefont {Portoles}},\ }\bibfield  {title} {\bibinfo
  {title} {{Kaon Decays in the Standard Model}},\ }\href
  {https://doi.org/10.1103/RevModPhys.84.399} {\bibfield  {journal} {\bibinfo
  {journal} {Rev. Mod. Phys.}\ }\textbf {\bibinfo {volume} {84}},\ \bibinfo
  {pages} {399} (\bibinfo {year} {2012})},\ \Eprint
  {https://arxiv.org/abs/1107.6001} {arXiv:1107.6001 [hep-ph]} \BibitemShut
  {NoStop}%
\end{thebibliography}%

\end{document}